\documentclass[final,leqno,showlabe]{siamltex}
\usepackage{amsmath}
\usepackage{amssymb}
\usepackage{cases}
\usepackage{enumerate}
\usepackage{graphicx}
\usepackage{doi}
\usepackage{cleveref}
\usepackage[usenames]{color}
\usepackage{wrapfig}
\usepackage{stmaryrd}
\usepackage{array} % <- Preamble
\usepackage[caption=false]{subfig} 
\SetSymbolFont{stmry}{bold}{U}{stmry}{m}{n}
\newtheorem{thm}{Theorem}[section]

\newcommand{\norm}[1]{\lVert#1\rVert}

\usepackage{multirow}

\newcommand{\be}{\begin{equation}}
\newcommand{\ee}{\end{equation}}
\newcommand{\ba}{\begin{array}}
\newcommand{\ea}{\end{array}}
\newcommand{\bea}{\begin{eqnarray}}
\newcommand{\eea}{\end{eqnarray}}

\newcommand{\dL}{{\rm d}{\mathcal{L}}}
\newcommand{\ds}{{\rm d}s}

\renewcommand{\vec}[1]{\mathbf{#1}}

\renewcommand{\b}[1]{\boldsymbol{#1}}

\title{A finite element method for Electrowetting on Dielectric}

\author{Quan Zhao\thanks{Department of Mathematics, National University of Singapore, 
Singapore 119076 (matzq@nus.edu.sg). 
}
\and Weiqing Ren\thanks{Corresponding author. 
Department of Mathematics, National University of
Singapore, Singapore, 119076 (matrw@nus.edu.sg). }
}

\date{}
\begin{document}

\maketitle
%%%%% Begin Abstract %%%%%%%%%%%

\begin{abstract}
We consider the problem of electrowetting on dielectric (EWoD).
The system involves the dynamics of a conducting droplet, which is  
immersed in another dielectric fluid, on a dielectric substrate
under an applied voltage.
The fluid dynamics is modeled by the two-phase incompressible 
Navier-Stokes equations with the standard interface conditions, 
the Navier slip condition on the substrate and a contact angle condition 
which relates the dynamic contact angle and the contact line velocity,
as well as the kinematic condition for the evolution of the interface.
The electric force acting on the fluid interface is modeled 
by the Maxwell's equations in the domain occupied by the 
dielectric fluid and the dielectric substrate.
We develop a numerical method for the model based on its weak form.  
This method combines the finite element method for 
the Navier-Stokes equations on a fixed bulk mesh 
with a parametric finite element method for the dynamics of the fluid interface, 
and the boundary integral method for the electric force 
along the fluid interface. 
Numerical examples are presented to demonstrate 
the accuracy and convergence of the numerical method, the effect of various physical 
parameters on the interface profile and other interesting phenomena such as the
transportation of droplet driven by applied non-uniform electric potential difference.

\end{abstract}
%%%%% end %%%%%%%%%%%

%%%%% Keywords %%%%%%%%%%%

\begin{keywords} Electrowetting, moving contact lines, two-phase flows, 
the finite element method
\end{keywords}

\begin{AMS}
74M15, 76D05, 76T10,  76M10, 76M15
\end{AMS}

\pagestyle{myheadings} \markboth{Q.~Zhao and W.~Ren}
{ A Finite Element Method for Electrowetting on Dielectric }

%=============================================Introduction===============================================
%=====================================================================================================
\section{Introduction}

\begin{figure}[!t]
\centering
\includegraphics[width=1.0\textwidth]{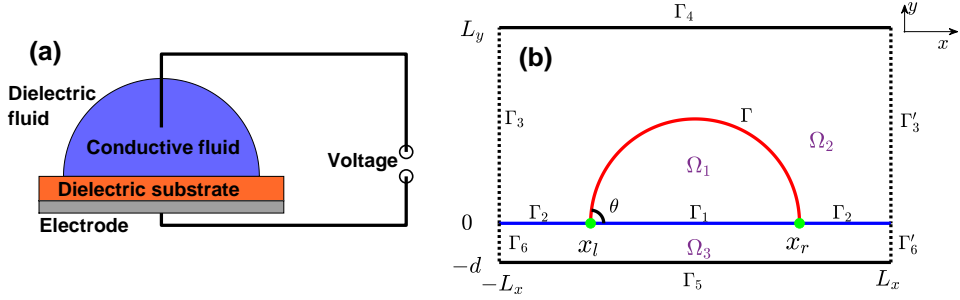}
\caption{(a) An illustration of the EWoD, where a conductive liquid droplet (shaded in blue) is deposited on a dielectric-coated  electrode (shaded in gray). (b) Geometric setup of the EWoD system with moving contact lines in two-phase flows in a bounded domain where $\Omega_1\cup\Omega_2=[-L_x,~L_x]\times[0,~L_y]$ and $\Omega_3=[-L_x,~L_x]\times[-d,~0]$.} 
\label{fig:model}
\end{figure}

% Electro-wetting, EWoD and applications
Since the pioneer work of Lippmann~\cite{Lippmann1875} on electro-capillarity,
it has been found that applied electric fields have a great effect 
on the wetting behavior of small charged droplets.
This phenomenon is referred to as electrowetting
and has received much attention in recent years \cite{Mugele05,Zhao13,Chen14}.
In the device of electro-wetting on dielectric (EWoD),
a dielectric film is placed on the substrate to separate the droplet 
and the electrode to avoid electrolytic decomposition \cite{Berge1993}
(see Fig. \ref{fig:model} for the set-up of EWoD).
EWoD has found many applications in various fields, such as adjustable 
lenses~\cite{Kuiper2004}, electronic displays \cite{Hayes2003}, 
lab-on-a-chip devices \cite{Pollack2002, Cho2003}, 
suppressing coffee strain effects \cite{Eral2011}, etc.

%% the static problem

The static problem of EWoD has been extensively studied in recent years, for example, in Refs. 
\cite{Kang2002,Buehrle2003, Shapiro2003, Bienia2006,
Mugele2007, Monnier2009, Scheid2009, Fontelos2012, Corson2014,Crowdy2015,Cui2019} and many others. 
These work has revealed the structure of the static interface profile. 
It was found that the electric force
does not contribute to the force balance at the contact line, therefore
the local static contact angle $\theta_Y$ still 
satisfies the Young-Dupr\'e equation 
\begin{equation}
\label{eqn:young}
\gamma\cos\theta_Y = \gamma_2 - \gamma_1,
\end{equation}
where $\gamma$, $\gamma_1$ and $\gamma_2$ are the surface tension coefficients
of the fluid-fluid and fluid-solid interfaces.
On the other hand, the divergent electric force incurs a large curvature 
and causes a significant deformation of the fluid interface in a small neighborhood of the contact line. 
The contact angle of the interface outside this small region, called
the apparent contact angle and denoted by $\theta_B$, is well characterized 
by the Lippmann equation~\cite{Quilliet2001,Buehrle2003,Cui2019}
\begin{equation}
\cos\theta_{B}=\cos\theta_Y + \frac{\epsilon\phi^2}{2\gamma\,d},
\label{eqn:Liapp}
\end{equation}
where $\phi$ is the applied voltage, $\epsilon$ and $d$ are the permittivity and thickness of the dielectric substrate, respectively.
Except in the extreme case of contact angle saturation,
this equation also matches experimental results quite well 
for different types of droplets and insulators, and wide range of $\phi$ and $d$ (see \cite{Vallet1999,Verheijen1999,Moon2002} for example).

In this work, we consider the dynamical problem of EWoD.  
Because of its importance in industrial applications, 
a lot of efforts have been devoted to this problem and 
some numerical methods have been proposed in recent years.
These include Lattice Boltzmann methods 
\cite{Clime2010, Li2009,Ruiz2019}, molecular dynamics simulations 
\cite{Daub2007,Kutana2006}, the level set method \cite{Walker2006,Guan2018},
the phase-field approach~\cite{Lu2007,Fontelos2012,Nochetto2014},
and others \cite{Cho2003, Nahar2015, Corson2016}, etc.
Here, we develop a finite element method for EWoD based on our earlier work on moving contact lines \cite{Zhao19}.

The model we use in this work for EWoD is based on the contact line model developed by Ren et al. \cite{Ren07, Ren10, Ren11d}. 
It contains the incompressible Navier-Stokes equations for the two-phase 
fluid dynamics, the Navier slip condition on the substrate 
and a dynamic contact angle condition at the contact line. 
On the fluid interface, besides the viscous stress and the curvature force,
the electric force also contributes to the force balance thus the interface conditions. We assume that the electric charging time is negligible compared 
to the time scale of the fluid motion, therefore model the electrostatic potential using the Maxwell's equation \cite{Cui2019}.

Based on the previous work for 
two-phase fluid dynamics \cite{Barrett15stable} 
and moving contact lines \cite{Zhao19}, we develop a finite element method for the EWoD model.
The method couples the finite element method for the incompressible 
Navier-Stokes equations and a semi-implicit parametric finite element method for the evolution of the fluid interface. We use unfitted mesh 
such that the discretization of the moving fluid interface is decoupled from the fixed bulk mesh. Besides, the electric force on the fluid interface is computed by using the boundary integral method. 
The numerical method obeys a similar energy law 
as the continuum model when the electric field is absent, which is a desired property for the numerical method.

This paper is organzied as follows. In section~\ref{sec:model}, 
we present the EWoD model and its dimensionless form. 
In section~\ref{sec:numerical}, we develop the numerical method based on a weak 
form of the continuum model, and present the full discretized scheme 
for the Navier-Stokes equations, the dynamics of the fluid interface, 
and the electric force on the fluid interface. 
In section~\ref{sec:num}, we present numerical examples
to demonstrate the convergence and accuracy of the numerical method, 
the effect of the various physical parameters on the interface profiles, as well as
the wetting dynamics driven by non-uniform electric potential. 
Finally, we draw the conclusion in section~\ref{sec:con}.

\section{The mathematical model}
\label{sec:model}

We consider the EWoD problem in the 2d space as shown in Fig.~\ref{fig:model}(a).  
In this setup,  a conductive liquid droplet (shaded in blue), is immersed 
in another dielectric fluid such as vapor or oil, and deposited on 
a thin dielectric substrate/insulator (shaded in orange), below which 
we place an electrode (shaded in gray). Between the electrode and the droplet, 
we apply a voltage difference, which generates electric fields 
and influences the wetting behavior of the charged droplet. 

The corresponding mathematical setup is shown in Fig.~\ref{fig:model}(b). 
We consider the system in a bounded domain and use the Cartesian coordinates, 
where the fluid-insulator interface is on the $x$-axis. For simplicity, 
we assume the periodic structure along $x$-direction for the system, 
and moreover, another electrode is placed on the top $\Gamma_4$. 
The domain is composed of three regions, the conducting droplet, 
the dielectric fluid, and the dielectric substrate, which are labeled as $\Omega_1$, $\Omega_2$, and $\Omega_3$, respectively. The fluid interface is denoted 
by the open curve $\Gamma$, which intersects with the flat insulator 
at two contact points, labeled as $x_l$ and $x_r$. $\Gamma_1$ and $\Gamma_2$ 
represent the corresponding fluid-insulator interfaces, 
and $\Gamma_5: y=-d$ is the insulator-electrode interface. 

Some relevant parameters are defined as follows:
$\rho_i$ and $\mu_i$ are the density and viscosity of fluid $i\ (i=1, 2)$,
respectively; $\epsilon_i$ is permittivity of the dielectric medium 
in $\Omega_i$ ($i=2,3$);  
$\gamma$ denotes the surface tension of the fluid interface $\Gamma$,
and $\gamma_i$ denotes the surface tension of the fluid-solid interface
$\Gamma_i\ (i=1,2)$. 
Furthermore, we denote by $\vec{n}$ the unit normal vector on $\Gamma$ 
pointing to fluid 2, and $\vec{n}_w=(0, -1)^T$ and $\vec{t}_w=(1,0)^T$
the unit normal and tangent vector on $\Gamma_1\cup\Gamma_2$ respectively.

\subsection{Governing equations for the  fluid dynamics} \label{subsec:f}
The mathematical model is an extension of the contact line model 
proposed by Ren et al. \cite{Ren07,Ren10, Ren11d} to include the contribution of electric field in EWoD. 
It consists of the two-phase incompressible Navier-Stokes equations for the fluid dynamics and the Maxwell's equation for the electrostatic potential.

First we consider the two-phase fluid dynamics in the domain 
$\Omega = \Omega_1(t)\cup\Omega_2(t)$. 
Let $\vec u(\vec x, t):\;\Omega\times[0,~T]\to\mathbb{R}^2$ 
be the fluid velocity  and $p(\vec x, t):\;\Omega\times[0, T]\to\mathbb{R}$ be the pressure. The dynamic of the system is governed by the standard 
incompressible Navier-Stokes equations in $\Omega_i(t)(i=1,2)$,
\begin{subequations}
\label{eqn:fluidynamic}
\begin{align}
\label{eqn:fluidynamic1}
\rho_i (\partial_t\vec u + \vec u\cdot\nabla\vec u) 
& = -\nabla p + \nabla\cdot\tau_d,\\ %[0.5em]
 \nabla\cdot\vec u & = 0,
 \label{fluidynamic2}
\end{align}
\end{subequations}
where $\tau_d=2\mu_iD(\vec u)$ is the viscous stress with 
$D(\vec u)=\frac{1}{2}(\nabla\vec u + (\nabla\vec u)^T)$.

On the fluid interface $\Gamma(t)$, we have the following conditions
\begin{subequations}\label{eqn:bdp1}
\begin{align}
\label{eqn:bd1a}
[\vec u]_1^2&=\vec 0,\\
\label{eqn:bd1b}
[p\mathbf{I}_2 - \tau_d]_1^2\cdot\vec n &
=\left( \gamma\kappa + %\frac{\sigma^2}{2\epsilon_2}
\frac{\epsilon_2}{2}|\nabla\Phi|^2
\right) \vec n,\\
\label{eqn:bd1c}
 v_n &= \left.\vec u\right|_{\Gamma(t)}\cdot\vec n,
\end{align}
\end{subequations}
where $[\cdot]_1^2$ denotes the jump from fluid 1 to fluid 2, 
$\mathbf{I}_2\in \mathbb{R}^{2\times2}$ is the identity matrix, 
$\kappa$ is the curvature of the fluid interface 
$\Gamma$, $\Phi$ is the electrostatic potential,  
and $v_n$ denotes the normal velocity of the fluid interface. 
Equation \eqref{eqn:bd1a} states the fluid velocity is continuous across the interface.
Equation \eqref{eqn:bd1b} states that the tangential stress is continuous across the interface and the normal stress jump is balanced by the
curvature force $\gamma\kappa\vec{n}$ together with the electric force $\frac{\epsilon_2}{2}|\nabla\Phi|^2\vec{n}$.
Equation \eqref{eqn:bd1c} is the kinetic condition 
for the interface, where the fluid interface evolves according to the local fluid velocity.

At the lower solid wall $\Gamma_i(t)\ (i=1,2)$, 
the fluid velocity satisfies the no-penetration condition 
and the Navier boundary condition 
\begin{subequations}\label{eqn:bdp2}
\begin{align}\label{eqn:bd2a}
\vec u\cdot\vec n_w &= 0,\\
\label{eqn:bd2b}
\vec t_w\cdot\tau_d\cdot\vec n_w &= -\beta_iu_s,
\end{align}
\end{subequations}
where $\beta_i\ (i=1,2)$ is the friction coefficient of fluid $i$ at the wall, and $u_s = \vec u\cdot\vec t_w$ is the slip velocity of the fluids. 

At the contact points $x_l$ and $x_r$, the dynamic contact angles, denoted by $\theta_d^l$ and $\theta_d^r$ respectively, depend on the contact line velocity \cite{Ren07,Ren10},
\begin{subequations}
\label{eqn:bdp3}
\begin{align}
\label{eqn:bd3da}
&\gamma(\cos\theta_d^l-\cos\theta_Y) = \beta^*\dot{x_l},\\
\label{eqn:bd3b}
&\gamma(\cos\theta_d^r-\cos\theta_Y)=-\beta^*\dot{x_r},
\end{align} 
\end{subequations}
where $\theta_Y$ is the equilibrium contact angle satisfying 
the Young's relation \eqref{eqn:young}, 
and $\beta^*$ is the friction coefficient of the fluid interface 
at the contact line. The contact line velocity is determined by 
the slip velocity of the fluids: $\dot{x}_{l,r}=\left.u_s\right|_{x=x_{l,r}}$. 
The contact angle condition \eqref{eqn:bdp3} is a force balance at the contact point:
the Young stress resulted from the deviation of the dynamic contact angle 
from the equilibrium angle is balanced by the friction force. 
Note that the electric force does not contribute to the force balance at the contact line.

Furthermore, we use the no-slip boundary condition on the upper wall $\Gamma_4$ 
and the periodic boundary conditions along $\Gamma_3$.

\subsection{Governing equations for the electrostatic field} \label{subsec:e}
The applied voltage induces electrostatic fields in the fluid region
$\Omega_2$ and the dielectric substrate $\Omega_3$, where 
the electric potential, denoted by $\Phi$, satisfies the Laplace equation,
\begin{equation}
\label{eqn:phi1}
\nabla^2\Phi = \dfrac{\partial^2\Phi}{\partial x^2} 
+ \dfrac{\partial^2\Phi}{\partial y^2}=0,\quad \vec{x}\in \Omega_2,\ \Omega_3.
\end{equation}
On the interface $\Gamma\cup\Gamma_1$ and at the top\slash bottom electrode
$\Gamma_4\cup\Gamma_5$, 
the electric potential satisfies the Dirichlet boundary conditions 
\begin{subequations}  \label{eqn:phi2}
\begin{align}
& \Phi = \phi,\quad{\rm on}\ \Gamma\cup\Gamma_1,\\
& \Phi = 0,\quad{\rm on}\ \Gamma_4\cup\Gamma_5,
\end{align}
\end{subequations}
where $\phi >0$ is the imposed voltage difference. 
Since there is no free charge density across the interface 
between the dielectric fluid and the dielectric substrate, we have 
\begin{equation}
\label{eqn:phi4}
\vec n_w\cdot \left[\epsilon \nabla\Phi\right]_2^3 =0,
\end{equation}
where $\left[\cdot\right]_2^3$ denotes the jump from $\Omega_2$ to
$\Omega_3$, $\epsilon=\epsilon_2$ in $\Omega_2$ and 
$\epsilon=\epsilon_3$ in $\Omega_3$. 
Furthermore, the periodic boundary condition is prescribed along $\Gamma_3\cup\Gamma_6$.

\subsection{Nondimensionlization}
\label{sec:dimensionless}
Next we present the above governing equations and the boundary/interface conditions
in their dimensionless form. By choosing $L$ and $U$ as the characteristic length 
and velocity respectively, we rescale the physical quantities as
\begin{align*}
&\hat{\rho}_i = \frac{\rho_i}{\rho_1},\quad \hat{\mu_i} = \frac{\mu_i}{\mu_1}, 
\quad\hat{\beta_i} = \frac{\beta_i}{\beta_1}, 
\quad \hat{\beta^*} = \frac{\beta^*}{\mu_1},
\quad\hat{\gamma}_i = \frac{\gamma_i}{\gamma},
\quad  \hat{\epsilon}_i = \frac{\epsilon_i}{\epsilon_3} ,\\
& \hat{\vec x} = \frac{\vec x}{L},\quad\hat{t} = \frac{Ut}{L},
\quad \hat{\vec u} = \frac{\vec u}{U},\quad \hat{p} = \frac{p}{\rho_1 U^2},\quad \hat{\kappa} = L\kappa,\quad \hat{\Phi} = \frac{\Phi}{\phi}.
\end{align*}
We define the Reynolds number $Re$, the Capillary number $Ca$, 
the slip length $l_s$, the Weber number $We$ and the parameter $\eta$ as
\begin{equation}
Re = \frac{\rho_1UL}{\mu_1},\quad Ca = \frac{\mu_1U}{\gamma}, 
\quad l_s = \frac{\mu_1}{\beta_1L},\quad We = Re\cdot Ca,
\quad  \eta = \frac{\epsilon_3\phi^2}{2\gamma L},
\end{equation}
where $\eta$ measures the relative strength of the electric force compared 
to the surface tension at the fluid interface.

For ease of presentation, we will drop the hats on the dimensionless parameters and variables. In the dimensionless form, the governing equations 
for the fluid dynamics in $\Omega_i\ (i=1,2)$ read
\begin{subequations}
\label{eqn:model12}
\begin{align}
\label{eqn:model1}
\rho_i(\partial_t\vec u + \vec u\cdot\nabla\vec u) +\nabla\cdot\vec T =0,\\
\label{eqn:model2}
\nabla\cdot\vec u  = 0,
\end{align}
\end{subequations}
where $\vec T = p\mathbf{I}_2-\frac{1}{Re}\tau_d$ is the stress tensor. 
These equations are supplemented with the following boundary\slash interface 
conditions:
\begin{itemize}
\item [(i)] The interface conditions on $\Gamma(t)$:
\begin{subequations} \label{eq:bd1}
\begin{align}
  \bigl[\vec u\bigr]^2_1 = \vec 0,\\
We \bigl[\vec T\bigr]_1^2\cdot\vec n = 
\left(\kappa + \epsilon_2\eta\left|\nabla\Phi\right|^2\right)\vec n,\\
\kappa = \partial_{ss}\vec X\cdot\vec n, \label{eq:cv}\\
v_n = \vec u|_{\Gamma(t)}\cdot\vec{n},
  \label{eq:k}
\end{align}
\end{subequations}
where $\vec X$ denotes the fluid interface, 
$s$ is the arc length parameter with $s=0$ being at the left contact point.
\item [(ii)] The condition for the dynamic contact angles:
\begin{subequations} \label{eqn:bd3}
\begin{align}
&\frac{1}{Ca}(\cos\theta_d^l - \cos\theta_Y)=\beta^*\dot{x}_l(t), \\
& \frac{1}{Ca}(\cos\theta_d^r - \cos\theta_Y)=-\beta^*\dot{x}_r(t).
\end{align}
\end{subequations}
\item [(iii)] The boundary conditions on $\Gamma_1(t)\cup\Gamma_2(t)$:
\begin{equation}
\label{eqn:bd2}
\vec u\cdot\vec n_w=0,\quad  l_s \vec t_w\cdot\tau_d\cdot\vec n_w =-\beta_i u_s.
\end{equation}
\item[(iv)] The no-slip condition on $\Gamma_4$
\begin{equation}
\label{eqn:bd5}
\vec u = \vec 0.
\end{equation}
\item[(v)] Periodic boundary conditions on $\Gamma_3$:
\begin{equation} \label{eqn:bd4}
\left.\vec u\right|_{x=-L_x} = \left.\vec u\right|_{x=L_x}, \quad
\left.\vec T\right|_{x=-L_x} = \left.\vec T\right|_{x=L_x}.
\end{equation}
\end{itemize}
The governing equations for the electric field potential $\Phi(x,t)$ 
in $\Omega_i(i=2,3)$ read
\begin{equation}
\label{eqn:dphi1}
\nabla^2\Phi = 0,
\end{equation}
subject to the following boundary conditions:
\begin{itemize}
\item [(i)] The Dirichlet Boundary conditions on $\Gamma$, $\Gamma_1$, 
$\Gamma_4$ and $\Gamma_5$:
\begin{subequations}
\label{eqn:dphi2}
\begin{align}
& \Phi = 1,\quad{\rm on}\ \Gamma(t)\cup\Gamma_1(t), \\
& \Phi = 0,\quad{\rm on}\ \Gamma_4,\ \Gamma_5.
\end{align}
\end{subequations}
\item[(ii)] The interface condition on $\Gamma_2(t)$:
\begin{equation}
\label{eqn:dphi3}
[\Phi]_2^3=0,\qquad\vec n_w\cdot \left[\epsilon\nabla\Phi\right]_2^3 =0.
\end{equation}
\item [(iii)] Periodic boundary conditions on $\Gamma_3\cup\Gamma_6$:
\begin{equation} \label{eqn:dphi4}
\left.\Phi\right|_{x=-L_x} = \left.\Phi\right|_{x=L_x}, \quad
 \left.\frac{\partial\Phi}{\partial\vec n}\right|_{x=-L_x} 
= -\left.\frac{\partial\Phi}{\partial\vec n}\right|_{x=L_x},
\end{equation}
where $\vec{n}$ is the unit outward normal to $\Omega_2$ on $\Gamma_3$. 
\end{itemize} 

\vspace{0.5cm}
Equations \eqref{eqn:model12} and \eqref{eqn:dphi1} together with the
boundary/interface conditions \eqref{eq:bd1}-\eqref{eqn:bd4} and
\eqref{eqn:dphi2}-\eqref{eqn:dphi4} form the complete model for the problem of EWoD. 
For this dynamical system, we define the total energy as
\begin{align}
W(t)&=\sum_{i=1,2}\int_{\Omega_i(t)}\frac{1}{2}\rho_i|\vec u|^2 \dL^2 
-\frac{\cos\theta_Y}{We}|\Gamma_1(t)| + \frac{1}{We}|\Gamma(t)|\nonumber\\
& \qquad -\frac{\eta}{We} \sum_{i=2,3}\int_{\Omega_i}
\epsilon_i |\nabla\Phi|^2 \dL^2,
\label{eqn:dimenenergy}
\end{align}
where $|\Gamma_1(t)|$ and $|\Gamma(t)|$ are the total arc length of $\Gamma_1(t)$ and $\Gamma(t)$, respectively.  
The four terms represents the kinetic energy of the fluids, 
the interfacial energy at the solid wall, the interfacial energy 
of the fluid interface and the electrical energy, respectively.
The dynamical system obeys the energy dissipation law
\begin{align}
\frac{{\rm d}}{{\rm d}t}W(t) &= -\sum_{i=1,2}\frac{1}{Re}
\int_{\Omega_i}2 \mu_i \|D(\vec u)\|_F^2 \dL^2
-\sum_{i=1,2}\frac{1}{Re\cdot l_s}\int_{\Gamma_i}\beta_i|u_s|^2 \ds\nonumber\\
& \qquad-\;\frac{\beta^{*}}{Re}(\dot{x_l}^2 + \dot{x_r}^2),
\label{eqn:dimensionenergydissipation2d}
\end{align}
where the three terms represent the viscous dissipation in the bulk fluid
with $\|\cdot\|$ being the Frobenius norm,
the dissipation at the solid wall due to the friction and the dissipation at the contact points, respectively. Details for the derivation of 
\eqref{eqn:dimensionenergydissipation2d} are provided in the Appendix~\ref{ap:energylaw}.

\section{The numerical method}\label{sec:numerical}

The numerical method consists of a finite element method for the fluid dynamics,
a parametric finite element method for the dynamics of the fluid interface, and 
the boundary integral method for the electric field. 
The numerical method is based on a weak formulation for the EWoD model, 
which we will present first.

\subsection{Weak formulation}\label{sec:weakform}
To propose the weak formulation for equations \eqref{eqn:model12}-\eqref{eqn:bd4}, 
we define the function space for the pressure and the velocity, respectively as
\begin{subequations}
\begin{align}
\mathbb{P}&:=\left\{\zeta\in L^2(\Omega): \;
\int_{\Omega}\zeta \dL^2=0\right\},\\
\label{eqn:USpace}
\mathbb{U}&:=\Bigl\{\boldsymbol{\omega}=(\omega_1,~\omega_2)^T\in 
\left(H^1(\Omega)\right)^2: \;\boldsymbol{\omega}\cdot\vec n_w =0\;
{\rm on}\;\Gamma_1\cup\Gamma_2,\;\boldsymbol{\omega}=\vec 0\;
{\rm on}\;\Gamma_4,\\
&\qquad\qquad\qquad \boldsymbol{\omega}(-L_x,y) =\boldsymbol{\omega}(L_x,y),
\;\forall \ 0\leq y\leq L_y \Bigr\},\nonumber
\end{align}
\end{subequations}
with the $L^2$-inner product on $\Omega=\Omega_1(t)\cup\Omega_2(t)$
\begin{equation}
(u, v):=\sum_{i=1,2}\int_{\Omega_i(t)} u v \,\dL^2,
\qquad\forall\ u,  v\in L^2(\Omega).
\end{equation}
We parameterize the fluid interface as 
$\vec X(\alpha, t)=(X(\alpha, t), Y(\alpha, t))^T$, 
where $\alpha\in {\mathbb{I}}=[0,1]$, and $\alpha=0, 1$ corresponding to the left and
and right contact points, respectively. 
We define the function spaces for the interface curvature and the interface as
\begin{subequations}
\begin{align}
& \mathbb{K} := L^2({\mathbb{I}})=\left\{\psi:\; 
    \mathbb{I}\rightarrow \mathbb{R}, \;\text{and} 
\int_\mathbb{I} |\psi(\alpha)|^2 |\partial_\alpha\vec X| {\rm d}\alpha <+\infty \right\},\\
& \mathbb{X}: = \Big\{ \b{g}=(g_1, g_2)^T\in (H^1(\mathbb{I}))^2:\; 
      g_2|_{\alpha=0, 1}=0\Big\},
\end{align}
\end{subequations}
equipped with the $L^2$-inner product on $\mathbb{I}$
\be
\big(u, v\big)_{\Gamma}:=
\int_{\mathbb{I}}u(\alpha)v(\alpha)\left|\partial_\alpha\vec{X}\right| 
{\rm d}\alpha,\qquad \forall\ u, v\in L^2(\mathbb{I}).
\ee

Taking the inner product of \eqref{eqn:model1} with 
$\boldsymbol{\omega}\in \mathbb{U}$ then  
using the boundary\slash interface conditions in 
\eqref{eq:bd1}-\eqref{eqn:bd4}
and $\nabla\cdot\vec u = 0$, we obtain \cite{Barrett15stable, Zhao19}
\begin{align}
&\Bigl(\rho [\partial_t\vec u + (\vec u\cdot\nabla)\vec u], \boldsymbol{\omega}\Bigr)
  \nonumber\\
=& \frac{1}{2}\left[\frac{{\rm d}}{{\rm d}t}\left(\rho\,\vec u, \boldsymbol{\omega}
\right) + \left(\rho\,\partial_t\vec u, \boldsymbol{\omega}\right)\right] 
+ \frac{1}{2}\Bigl(\rho, [(\vec u\cdot\nabla)\vec u]
\cdot\boldsymbol{\omega} - [(\vec u\cdot\nabla)\boldsymbol{\omega}]\cdot\vec u\Bigr),
\label{eqn:weak11}
\end{align}
where $\rho = \rho_1\chi_{_{\Omega_1(t)}} + \rho_2\chi_{_{\Omega_2(t)}}$ with
$\chi$ being the characteristic function. 
With the special treatment of the inertia term in \eqref{eqn:weak11},
the numerical scheme enjoys the discrete stability for the fluid kinetic energy in the absence of electric field.
This will be discussed later in section \ref{subsec:p}. 
For the viscous term, integrating by parts then 
applying the boundary\slash interface conditions yields
\begin{align}
\label{eqn:weak12}
\Bigl(\nabla\cdot\vec T,\, \boldsymbol{\omega}\Bigr)&
=-\Bigl(p,\, \nabla\cdot\boldsymbol{\omega}\Bigr) 
+ \frac{2}{Re}\Bigl(\mu D(\vec u),\,D(\boldsymbol{\omega})\Bigr)
 -\Bigl([\vec T]_1^2\cdot\vec n,\, \boldsymbol{\omega}\Bigr)_{\Gamma} \nonumber\\ 
&\qquad\qquad\qquad
+ \Bigl(\vec T\cdot\vec n_w,\, \boldsymbol{\omega}\Bigr)_{\Gamma_1\cup\Gamma_2}
\nonumber\\
&= -\Bigl(p,~\nabla\cdot\boldsymbol{\omega}\Bigr) 
+ \frac{2}{Re}\Bigl(\mu D(\vec u),~D(\boldsymbol{\omega})\Bigr) 
- \frac{1}{We}\Bigl(\kappa + \epsilon_2\eta|\nabla\Phi|^2,
~\vec n\cdot\boldsymbol{\omega} \Bigr)_{\Gamma} \nonumber\\ 
&\qquad\qquad\qquad
+ \frac{1}{Re\cdot l_s}\Bigl(\beta\,u_s,~\omega_s\Bigr)_{\Gamma_1\cup\Gamma_2}, 
\end{align}
where $\mu =\mu_1\chi_{_{\Omega_1(t)}} + \mu_2\chi_{_{\Omega_2(t)}}$, 
$\beta = \beta_1\chi_{_{\Gamma_1(t)}} + \beta_2\chi_{_{\Gamma_2(t)}}$, 
and $\omega_s = \boldsymbol{\omega}\cdot\vec t_w$.

Equation \eqref{eq:cv} for the curvature can be rewritten as 
$\kappa\vec n = \partial_{ss}\vec X$. Taking the inner product 
of it with $\b{g}=(g_1,~g_2)^T\in \mathbb{X}$ 
on $\Gamma(t)$ then applying integration by parts yields
\begin{align}
0& 
=\Bigl(\kappa, ~\vec n\cdot\b{g}\Bigr)_{\Gamma}
+\Bigl(\partial_s\vec X, ~\partial_s\b g\Bigr)_{\Gamma} 
- (\partial_s\vec X\cdot\b g)\Big|_{\alpha=0}^{\alpha=1}\nonumber\\
&=\Bigl(\kappa,~ \vec n\cdot\b g\Bigr)_{\Gamma}
 +\Bigl(\partial_s\vec X, ~\partial_s\b g\Bigr)_{\Gamma} 
-(g_1 \partial_sX)\Big|_{\alpha=0}^{\alpha=1}\nonumber\\
&=\Bigl(\kappa, ~\vec n\cdot\b g\Bigr)_{\Gamma}
+\Bigl(\partial_s\vec X, ~\partial_s\b g\Bigr)_{\Gamma} 
+ \beta^*\,Ca\left[\dot{x}_l g_1(0) + \dot{x}_r g_1(1)\right]\nonumber\\
&\qquad\qquad\qquad\qquad\qquad 
 - \cos\theta_Y \left[g_1(1)-g_1(0)\right],
\label{eqn:curvaturefor}
\end{align}
where we have used the fact that $\partial_s X|_{\alpha=0} = \cos\theta_d^l$, 
$\partial_s X|_{\alpha=1}=\cos\theta_d^r$, and 
the dynamic contact angle conditions in \eqref{eqn:bd3}.

Combining these results, we obtain the weak formulation for the dynamic system
\eqref{eqn:model12}-\eqref{eqn:bd4} as follows: Given the initial fluid velocity 
$\vec u_0$ and the interface $\vec X_0(\alpha)$, find the fluid velocity 
$\vec u(\cdot,~t)\in \mathbb{U}$, the pressure $p(\cdot,~t)\in\mathbb{P}$, 
the fluid interface $\vec X(\cdot,~t)\in \mathbb{X}$, 
and the curvature $\kappa(\cdot,~t)\in \mathbb{K}$ 
such that
\begin{subequations}
\begin{align}
&\frac{1}{2} \left[\frac{{\rm d}}{{\rm d}t}\Bigl(\rho\,\vec u,~\boldsymbol{\omega}
\Bigr)+\Bigl(\rho\,\partial_t\vec u,~\boldsymbol{\omega}\Bigr)
+ \Bigl(\rho\,(\vec u\cdot\nabla)\vec u,~\boldsymbol{\omega}\Bigr)
-\Bigl(\rho\,(\vec u\cdot\nabla)\boldsymbol{\omega},~\vec u\Bigr)\right] \nonumber\\
&\quad+\frac{2}{Re} \Bigl(\mu D(\vec u),~D(\boldsymbol{\omega})\Bigr)
-\Bigl(p,~\nabla\cdot\boldsymbol{\omega}\Bigr) 
-\frac{1}{We}\Bigl(\kappa\,\vec n,~\boldsymbol{\omega}\Bigr)_{\Gamma} \nonumber\\
&\quad- \frac{\epsilon_2\eta}{We}\Bigl(|\nabla\Phi|^2 \vec n,
~\boldsymbol{\omega}\Bigr)_{\Gamma} + \frac{1}{Re\cdot l_s}
\Bigl(\beta u_s,~\omega_s\Bigr)_{\Gamma_1\cup\Gamma_2} =0,
\quad\forall\, \boldsymbol{\omega}\in \mathbb{U},
      \label{eqn:weak1}\\[0.6em]
& \qquad\hspace{2cm} \Bigl(\nabla\cdot\vec u,~\zeta\Bigr) =0, 
\quad\forall\, \zeta\in\mathbb{P},
      \label{eqn:weak2a}\\[0.6em]
                        \label{eqn:weak2}
&\qquad\quad \Bigl(\vec n\cdot\partial_t\vec X,~\psi\Bigr)_{\Gamma} 
- \Bigl(\vec u\cdot\vec n,~\psi\Bigr)_{\Gamma}=0,\quad\forall\, 
\psi\in \mathbb{K},\\[0.6em]
&\Bigl(\kappa,~\vec n\cdot\boldsymbol{g}\Bigr)_{\Gamma}
+\Bigl(\partial_s\vec X,~\partial_s\boldsymbol{g}\Bigr)_{\Gamma}
+\beta^* Ca\Bigl[\dot{x}_l\,g_1(0) + \dot{x}_r\,g_1(1)\Bigr]\nonumber\\
&\qquad\qquad\quad- \cos\theta_Y [g_1(1) - g_1(0)] = 0,\quad\forall\, 
\boldsymbol{g}\in \mathbb{X}.
\label{eqn:weak3}
\end{align}
\end{subequations}
Eq. \eqref{eqn:weak1} is obtained from Eq. \eqref{eqn:weak11} and Eq. \eqref{eqn:weak12}. Eq. \eqref{eqn:weak2a} results from the incompressibility condition \eqref{eqn:model2}. Eq. \eqref{eqn:weak2} is obtained from 
the kinetic condition \eqref{eq:k}. 
Eq. \eqref{eqn:weak3} is obtained from \eqref{eqn:curvaturefor}. 

The above system \eqref{eqn:weak1}-\eqref{eqn:weak3} is an extension 
of the weak formulation introduced in Ref.~\cite{Barrett15stable} for two-phase
flows and Ref.~\cite{Zhao19} for moving contact lines. 
In the current problem for EWoD, we have the additional term for the electric force 
in \eqref{eqn:weak1}. The electric force
is obtained from solving \eqref{eqn:dphi1}-\eqref{eqn:dphi4}.

\subsection{Finite element discretization}\label{sec:FEM}

We solve equations \eqref{eqn:weak1}-\eqref{eqn:weak3} for $\vec{u}$ and $p$
on the fluid domain $\Omega$ and for $\vec{X}$ and $\kappa$ on the reference domain
$\mathbb{I}$ using the finite element method on unfitted meshes.
To that end, we uniformly partition the time domain as 
$[0, T]=\cup_{m=1}^M [t_{m-1}, t_m]$ with $t_m = m\tau$, $\tau=T/M$,  and the reference domain as
$\mathbb{I}=\cup_{j=1}^{J_{\Gamma}}\mathbb{I}_j$ 
with $\mathbb{I}_j=[\alpha_{j-1}, \alpha_j]$, $\alpha_j = jh$ and $h=1/J_{\Gamma}$. 
We approximate the function spaces $\mathbb{K}$ and $\mathbb{X}$ by
the finite element spaces
\begin{subequations}
\begin{align}
&\mathbb{K}^h:=\Big\{\psi\in C(\mathbb{I}): \,\psi|_{\mathbb{I}_j}\in 
\mathcal{P}_1(\mathbb{I}_j),\quad \forall\, j = 1,2,\cdots, J_{\Gamma}\Big\},\\
& \mathbb{X}^h:=\Big\{ \b{g}=(g_1, g_2)^T \in (C(\mathbb{I}))^2: \,
\b{g}|_{\mathbb{I}_j}\in (\mathcal{P}_1(\mathbb{I}_j))^2,  \quad
\forall\, j = 1,2,\cdots, J_{\Gamma}; \\
& \qquad \qquad\qquad g_2|_{\alpha=0, 1} =0\Big\}, \notag 
\end{align}
\end{subequations}
where $\mathcal{P}_1(\mathbb{I}_j)$ denotes the space of the polynomials of degree 
at most 1 over $\mathbb{I}_j$. Denote by $\Gamma^m:=\vec X^m(\cdot)\in\mathbb{X}^h$
the numerical approximation to the fluid interface $\Gamma(t)$ 
at $t=t_m$. Then $\Gamma^m\ (0\leq m\leq M)$ are polygonal curves consisting of ordered line segments.
The unit tangential vector $\vec{t}^m$ and normal vector 
$\vec n^m$ of $\Gamma^m$ are constant vectors on each interval 
$\mathbb{I}_j$ with possible jumps at the nodes $\alpha_j$, 
and they can be easily computed as
\begin{equation}
\vec{t}_j^m:=\vec{t}^m|_{\mathbb{I}_j}=\frac{\vec h_j^m}{|\vec h_j^m|},\quad
\vec{n}_j^m:=\vec{n}^m|_{\mathbb{I}_j}
=({\vec{t}}^m_j)^{\perp}, \quad 
1\leq j\leq J_{_\Gamma}\label{eqn:semitannorm}
\end{equation}
where
$\vec h_j^m:=\vec X^m(\alpha_j)-\vec X^m(\alpha_{j-1})$, and
 $(\cdot)^\perp$ denotes the counter-clockwise rotation by $90$ degrees.

Let $\mathcal{T}^h=\left\{ \bar{o}_j\right\}_{j=1}^N$ 
be a triangulation of the bounded domain $\Omega$, and 
\begin{subequations}
\begin{align}
& S_k^h:=\left\{\varphi\in C(\bar{\Omega}):\; 
\varphi|_{o_j}\in \mathcal{P}_k(o_j),\;\forall\ j=1,2,\cdots,N\right\}, \\
& S_0^h:=\{\varphi\in L^2(\Omega):\; \varphi|_{o_j}\in 
\mathcal{P}_0(o_j),\; \forall j=1,2,\cdots,N\},
\end{align}
\end{subequations}
where $k\in \mathbb{N}^{+}$, and $\mathcal{P}_k(o_j)$ denotes the space of polynomials of degree $k$ on $o_j$. 
Let $\mathbb{U}^h$ and $\mathbb{P}^h$ denote the finite element spaces for the numerical solutions for the velocity and pressure, respectively. 
In this work, we choose them as
\begin{equation}
\label{eqn:spaceUP}
\Bigl(\mathbb{U}^h,~\mathbb{P}^h\Bigr)
=\Bigl(\left(S_2^h\right)^2\cap\mathbb{U},~
\left(S_1^h+S_0^h\right) \cap\mathbb{P}\Bigr),
\end{equation}
which satisfies the inf-sup stability condition \cite{Agnese16,Zhao19}
\begin{equation}
\inf_{\varphi\in  \mathbb{P}^h} 
\sup_{\vec 0\neq\boldsymbol{\omega}\in \mathbb{U}^h}
\frac{\left(\varphi, \nabla\cdot\boldsymbol{\omega}\right)}
{\norm{\varphi}_0\norm{\boldsymbol{\omega}}_1}\geq c>0,
\label{eqn:LBB}
\end{equation}
where $\norm{\cdot}_0$ and $\norm{\cdot}_1$ denote the $L^2$ and $H^1$-norm 
on $\Omega$ respectively, and $c$ is a constant.

\begin{figure}[!t]
\centering
\includegraphics[width=0.99\textwidth]{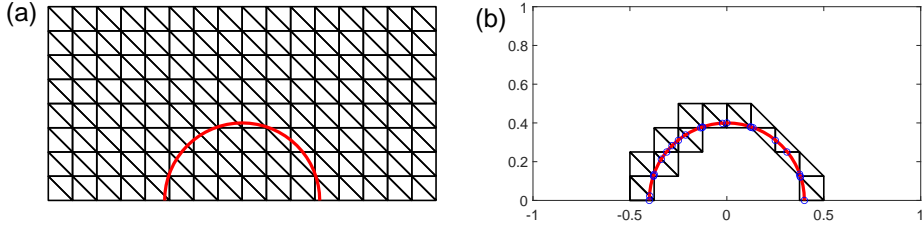}
\caption{(a) Illustration of the discretization of the fluid domain 
$\Omega$ and the fluid interface $\Gamma^m$ (red curve) at $t=t_m$. 
The fluid interface 
divides $\Omega$ into two sub-domains: $\Omega_1^m$ being the region
enclosed by $\Gamma^m$ and the lower wall, $\Omega_2^m$ being the region
outside. (b) Intersection of the interface $\Gamma^m$ with
the bulk mesh. The elements of intersection form $\mathcal{T}_f^m$.}
\label{fig:mesh}
\end{figure}

In the simulation, the partition of the reference domain $\mathbb{I}$ 
for the interface and the triangulation $\mathcal{T}^h$ 
of the fluid domain $\Omega$ are both fixed in time.
As a result, the triangulation of $\Omega$ is decoupled from the discretization
of the interface $\Gamma^m$, thus may not fit the interface,
i.e. the line segments comprising of $\Gamma^m$ are in general not the boundaries of the elements in $\mathcal{T}^h$. 
At $t=t^m$, the interface $\Gamma^m$ divides $\Omega$ into two sub-domains, $\Omega_1^m$ and $\Omega_2^m$.
Correspondingly, we split $\mathcal{T}^h$ 
into three disjoint subsets: $\mathcal{T}_1^m$ being the set of elements 
in $\Omega_1^m$,  $\mathcal{T}_2^m$ being the set of elements in $\Omega_2^m$,
and $\mathcal{T}_f^m$ being the set of elements that intersect with the interface
(see Fig. \ref{fig:mesh} for an illustration). 
The splitting can be easily done by using 
a recursive algorithm as follows:
\begin{itemize}
\item [(1)] First form $\mathcal{T}^m_f$ by locating all elements that intersect with $\Gamma^m$.
\item [(2)] Locate one element $o_{j*}$ in $\Omega_1^m$, 
for example, the one that contains the point with coordinate 
$(\frac{1}{2}(x_l^m+x_r^m),~0)$, and set $\mathcal{T}_1^m = \left\{o_{j*}\right\}$.
\item [(3)] Check all neighbours of the elements in $\mathcal{T}_1^m$. 
If a neighbor is not in $\mathcal{T}^m_f$, then add it to $\mathcal{T}_1^m$.
\end{itemize}
The last step is repeated until no element can be added to $\mathcal{T}_1^m$.
This gives the set $\mathcal{T}_1^m$. The rest of the elements not 
belonging to $\mathcal{T}_1^m\cup\mathcal{T}_f^m$ form the set $\mathcal{T}_2^m$.

Denote by $\rho^m$, $\mu^m$ and $\beta^m$ the numerical approximations of
the density $\rho(\cdot, t)$, the viscosity $\mu(\cdot, t)$ and the friction
coefficient $\beta(\cdot, t)$ at $t=t_m$, respectively. 
We define $\rho^m$ and $\mu^m$ as
\begin{equation}
\label{eqn:discreterhovis}
\rho^m|_{o_j}:=\left\{
\begin{array}{ll}
\rho_1, & \text{if}\ o_j\in \mathcal{T}_1^m, \vspace{0.15cm}\\
\rho_2, & \text{if}\ o_j\in\mathcal{T}_2^m,\vspace{0.15cm}\\
\frac{1}{2}(\rho_1+\rho_2), &\text{if}\ o_j\in \mathcal{T}_{f}^m,
\end{array}\right.\qquad
\mu^m|_{o_j}:=\left\{\begin{array}{ll}
\mu_1, &\text{if}\ o_j\in \mathcal{T}_1^m,\vspace{0.15cm}\\
\mu_2,&\text{if}\ o_j\in\mathcal{T}_2^m, \vspace{0.15cm}\\
\frac{1}{2}(\mu_1+\mu_2), & \text{if}\ o_j\in \mathcal{T}_{f}^m,
\end{array}\right.\nonumber
\end{equation}
where $1\leq j\leq N$, $0\leq m\leq M$. 
Similarly, we denote by $\Gamma_1^m$ and $\Gamma_2^m$ 
the boundary of $\Omega_1^m$ and $\Omega_2^m$ at the lower wall respectively, and
define $\beta^m$ at the lower wall as
\begin{equation}
\beta^m|_{\partial o_j}:=\left\{
\begin{array}{ll}
\beta_1, & \text{if}\ \partial o_j\subset\Gamma_1^m,\vspace{0.15cm}\\
\beta_2,& \text{if}\ \partial o_j \subset\Gamma_2^m,\vspace{0.15cm}\\
\frac{1}{2}(\beta_1+\beta_2), &\text{if}\  x_l^m\in \partial o_j\;
{\rm or}\; x_r^m\in \partial o_j,
\end{array}\right.
\end{equation}
where $\partial o_j$ is the boundary of $o_j$ at the lower wall, 
and $x_l^{m}=X^{m}|_{\alpha=0}$ and $x_r^m=X^m|_{\alpha=1}$
 are the two contact line points.

The finite element method is given as follows. First we partition the time domain
$[0, T]$, the reference domain $\mathbb{I}$ for the interface 
and the fluid domain $\Omega$ as described above. 
Let $\Gamma^0:=\vec X^0(\cdot)\in\mathbb{X}^h$ and $\vec u^0\in \mathbb{U}^h$
be the initial interface and velocity field, respectively.  
For $m\geq 0$, we compute $\vec u^{m+1}\in \mathbb{U}^h$, 
$p^{m+1}\in\mathbb{P}^h$, $\vec X^{m+1}\in \mathbb{X}^h$, 
and $\kappa^{m+1}\in \mathbb{K}^h$ by solving the linear system
\begin{subequations}
\begin{align}\label{eqn:full1}
&\frac{1}{2}\Bigl[\Bigl(\frac{\rho^m\vec u^{m+1}-\rho^{m-1}
\vec u^m}{\tau}, \boldsymbol{\omega}^h\Bigr)
+\Bigl(\rho^{m-1}\frac{\vec u^{m+1}-\vec u^m}
{\tau}, \boldsymbol{\omega}^h\Bigr)    \nonumber\\
& +\Bigl(\rho^m(\vec u^m\cdot\nabla)\vec u^{m+1}, \boldsymbol{\omega}^h\Bigr)
-\Bigl(\rho^m(\vec u^m\cdot\nabla)\boldsymbol{\omega}^h,~\vec u^{m+1}
\Bigr)\Bigr]- \Bigl(p^{m+1},~\nabla\cdot\boldsymbol{\omega}^h\Bigr)  \nonumber\\ 
&  +\frac{2}{Re}\Bigl(\mu^m D(\vec u^{m+1}),~D(\boldsymbol{\omega}^h)\Bigr) 
-\frac{1}{We}\Bigl(\kappa^{m+1}\vec n^m,~\boldsymbol{\omega}^h
\Bigr)_{\Gamma^m}  \nonumber \\  
& -\frac{\epsilon_2\eta}{We}
\Big(|\nabla\Phi|^2\,\vec n^m,~\boldsymbol{\omega}^h\Big)_{\Gamma^m} 
+ \frac{1}{Re\cdot l_s}\Bigr(\beta^m u_s^{m+1},
~\omega_s^h\Bigr)_{\Gamma_1^m\cup\Gamma_2^m}=0, \nonumber\\
& \qquad \hspace{7.4cm}
\forall\, \boldsymbol{\omega}^h
\in \mathbb{U}^h,  \\[0.7em]
\label{eqn:full2a}
& \qquad\qquad\qquad \Bigl(\nabla\cdot\vec u^{m+1},~\zeta^h\Bigr) =0,
\quad \forall\, \zeta^h\in\mathbb{P}^h,  \\[0.7em]
\label{eqn:full2}
&\Bigl(\frac{\vec X^{m+1}-\vec X^m}{\tau}\cdot\vec n^m,
~\psi^h\Bigr)_{\Gamma^m}^h - \Bigl(\vec u^{m+1}\cdot\vec n^m,~\psi^h\Bigr)
_{\Gamma^m}=0,\quad\forall\, \psi^h\in \mathbb{K}^h,\\[0.7em]
&\Bigl(\kappa^{m+1}\,\vec n^m,~\boldsymbol{g}^h\Bigr)_{\Gamma^m}^h
+\Bigl(\partial_s\vec X^{m+1},~\partial_s\boldsymbol{g}^h\Bigr)_{\Gamma^m}^h
 -\cos\theta_Y \left[g_1^h(1) - g_1^h(0)\right]\nonumber\\
&\quad 
+\frac{\beta^* Ca}{\tau}\left[\left(x^{m+1}_r- x^m_r\right)g_1^h(1)
+\left(x^{m+1}_l-x_l^m\right) g_1^h(0)\right]=0,
\quad\forall\,\boldsymbol{g}^h\in \mathbb{X}^h.
\label{eqn:full3}
\end{align}
\end{subequations}
Here, $\b{g}^h = (g_1^h,~g_2^h)^T$, $\omega_s^h=\omega^h\cdot \vec{t}_w$, 
$u_s^{m+1} = \vec{u}^{m+1}\cdot\vec{t}_w$, and $x_l^m = \vec{X}^m|_{\alpha=0}$,
$x_r^m = \vec{X}^m|_{\alpha=1}$. The 
 electric force
$\epsilon_2\eta|\nabla\Phi|^2$ in \eqref{eqn:full1} is computed by solving
equations \eqref{eqn:dphi1}-\eqref{eqn:dphi4} in the domain $\Omega_2^m\cup\Omega_3$;
its computation will be presented in section~\ref{sec:bem}. 
In \eqref{eqn:full3}, the derivative $\partial_s\vec{X}^{m+1}$ is taken
with respect to the arc length of $\Gamma^m$: 
$\partial_s\vec{X}^{m+1} = \frac{1}{|\partial_\alpha\vec{X}^m|}
\partial_\alpha\vec{X}^{m+1}$, and similarly for $\partial_s\b{g}^h$.
At the first time step, we set $\rho^{-1}=\rho^0$.

In the numerical method, the inner product $(\cdot, \cdot)_{\Gamma(t^m)}$ is 
approximated by using either the trapezoidal rule, denoted by 
$(\cdot, \cdot)_{\Gamma^m}^h$, or the Simpson's rule, denoted by 
$(\cdot, \cdot)_{\Gamma^m}$.  Since we are using unfitted mesh, 
the interface $\Gamma^m$ intersects with the bulk mesh. 
Denote by $\left\{\alpha'_j\right\}_{j=1}^{J'_\Gamma}$ the set, in ascending order, of
both the $\alpha$-values of the intersecting points and the mesh points 
of the reference interval $\mathbb{I}$, 
$\alpha_j=j/J_\Gamma\ (j=0, \cdots, J_{\Gamma})$.
Then the inner products involving interface and bulk quantities 
are approximated by the Simpson's rule on the mesh 
$\left\{\alpha'_j\right\}_{j=1}^{J'_\Gamma}$,
\begin{equation}
(u,v)_{_{\Gamma^m}}\!= \frac{1}{6}\sum_{j=1}^{J'_\Gamma}
\Big|\vec{X}^m(\alpha'_j)-\vec{X}^m(\alpha'_{j-1})\Big|
\Big[(u\cdot v)(\alpha'^+_{j-1}) +4 (u\cdot v)(\alpha'_{j-\frac{1}{2}})
+ (u\cdot v)(\alpha'^-_{j})\Big],
\end{equation}
and the inner products involving only quantities on the interface are simply 
approximated by the trapezoidal rule on the mesh 
$\left\{\alpha_j\right\}_{j=1}^{J_\Gamma}$,
\begin{align}
\label{eqn:massnorm}
&\big(u, v\big)_{\Gamma^m}^h =\frac{1}{2}\sum_{j=1}^{J_\Gamma} 
\Big|\vec{X}^m(\alpha_j)-\vec{X}^m(\alpha_{j-1})\Big|
\Big[\big(u\cdot v\big)(\alpha_{j-1}^+)
+\big(u\cdot v\big)(\alpha_j^-)\Big],
\end{align}
where $\alpha'_{j-\frac{1}{2}}=\frac{1}{2}\left(\alpha'_{j-1} + \alpha'_{j}\right)$, $u(\alpha'^\pm_j)$ are the one-sided limits of $u$ at $\alpha'_j$, and
similarly for $u(\alpha_j^\pm)$.

\begin{figure}[!t]
\centering
\includegraphics[width=0.7\textwidth]{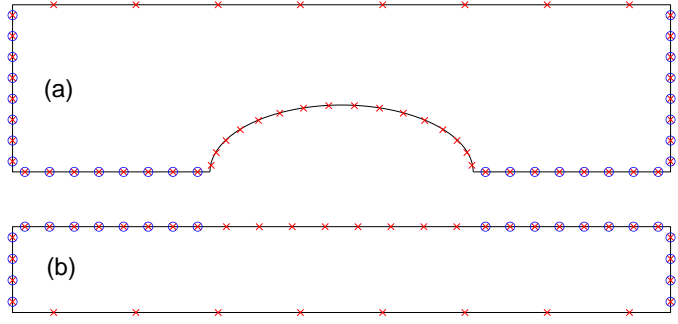}
\caption{The boundary integral method is applied to the domain $D_1=\Omega_2^m$ (upper panel) and the domain $D_2=\Omega_3$ (lower panel) separately.
The electrostatic potential $\Phi$ and its directional derivative $\Psi$
are evaluated at the discrete points indicated by $"\circ"$ and $"\times"$ respectively along the boundaries. }
\label{fig:bim}
\end{figure}

\subsection{Computation of the electric force}
\label{sec:bem}

The electric force on the fluid interface is computed by solving Eqs.~\eqref{eqn:dphi1}-\eqref{eqn:dphi4} using the boundary integral method.
Without loss of generality, we assume that $\Phi$ satisfies the Laplace equation 
on a domain $\mathcal{D}$ with boundary $\Sigma$. 
Denote the directional derivative of $\Phi$ in the outward normal direction
of the boundary by $\Psi = \vec{n}\cdot\nabla\Phi$, where $\vec{n}$ is the outward unit normal vector on $\Sigma$. 
For any point $\vec p$ lies on the smooth part of $\Sigma$, we have the boundary integral equation
\begin{equation}
\label{eqn:boudIntegral}
\frac{1}{2}\Phi(\vec p)=\int_{\Sigma}\left[\Phi(\vec q)
\frac{\partial G(\vec p, \vec q)}{\partial\vec{n}(\vec q)}
- \Psi(\vec q) G(\vec p, \vec q)\right] \ds(\vec q),
\end{equation}
where $G(\vec p, \vec q)=\frac{1}{2\pi}\ln|\vec p - \vec q|$ is the Green function for the Laplace equation
in $\mathbb{R}^2$, and $\frac{\partial G(\vec p, \vec q)}{\partial \vec{n}(\vec q)}
=\vec{n}\cdot\nabla_{\vec{q}} G(\vec p, \vec q)$.

Eq. \eqref{eqn:boudIntegral} can be used to compute $\Phi$ and/or its derivative
$\Psi$ on the boundary $\Gamma$. To that end, we partition $\Sigma$ then approximate
it by a collection of line segments: $\Sigma\simeq \cup_{j=1}^M \Sigma^{(j)}$.
We further approximate $\Phi(\vec p)$ and $\Psi(\vec p)$ by constants 
on each line segment:  $\Psi\simeq \Phi_j$ and $\Psi\simeq \Psi_j$
on $\Sigma^{(j)}$, $1\leq j\leq M$. Denote by $\vec p_j$ the mid-point of the line segment $\Sigma^{(j)}$. It lies on a smooth part of the approximation boundary $\Sigma^{(j)}$. Then we can apply Eq. \eqref{eqn:boudIntegral}
to $\vec{p} = \vec{p_j}$ and obtain
\begin{equation}
\label{eqn:disbem}
\frac{1}{2}\Phi_j = \sum_{k=1}^{M}\Bigl[A_{jk}\Phi_k - 
B_{jk}\Psi_k\Bigr],\quad j=1,\cdots, M,
\end{equation}
where
\begin{equation}
\label{eqn:disbem2}
A_{jk}=\int_{\Sigma^{(k)}}\frac{\partial G(\vec p_j, \vec q)}
{\partial\vec{n}(\vec q)}\ds(\vec q),\qquad
B_{jk}=\int_{\Sigma^{(k)}}G(\vec p_j, \vec q)\ds(\vec q).
\end{equation}

For the current problem, we apply the boundary integral method to the domain $D_1=\Omega_2^m$ and the domain $D_2=\Omega_3$ separately. 
 A discretization of the boundary of the two domains
is shown in Fig. \ref{fig:bim}. 
Eq. \eqref{eqn:disbem} is applied at each discrete point.
These equations, together with the prescribed Dirichlet boundary conditions
$\Phi|_{\Gamma^m}=\Phi|_{\Gamma_1}=1$ and  
$\Phi|_{\Gamma_4}=\Phi|_{\Gamma_5}=0$,
the periodic boundary conditions 
$\Phi|_{\Gamma_3}=\Phi|_{\Gamma'_3}$, 
$\Psi|_{\Gamma_3}=-\Psi|_{\Gamma'_3}$,
$\Phi|_{\Gamma_6}=\Phi|_{\Gamma'_6}$ and 
$\Psi|_{\Gamma_6}=-\Psi|_{\Gamma'_6}$,
as well as the interface conditions
$\left[\Phi\right]_2^3=0$ and $\left[\epsilon\Psi\right]_2^3=0$ on $\Gamma_2^m$ form
a system of linear algebraic equations for $\Phi$ and $\Psi$ at the discrete points. After the linear system is solved, $\Psi$ is used to compute the electric force:
$|\nabla\Phi|^2 = \Psi^2$ on $\Gamma^m$.

\subsection{Properties of the numerical method}  \label{subsec:p}

For the numerical method \eqref{eqn:full1}-\eqref{eqn:full3}, we can show that it yields a unique solution. Furthermore, in the special case 
when the electric force is absent, the numerical method is unconditionally energy stable. We make the following assumptions on  the
interface $\Gamma^m$, $\forall\, 0\le m\le M$, 
\begin{itemize}
\item[i)] the interface $\Gamma^m$ does not intersect with itself;
\item[ii)] the parameterization is such that $|\partial_\alpha\vec{X}^m|>0$, and
\item[iii)] the first and last segments of $\Gamma^m$ are not parallel to the $x$-axis.
\end{itemize}
These assumptions particularly imply that the mesh points on $\Gamma^m$ do not merge, and the dynamic contact angle is not 0 or $\pi$.

\vspace{0.15cm}
\begin{thm}[Existence and Uniqueness]
\label{th:wellposed}
Let the finite element spaces $(\mathbb{U}^h, \mathbb{P}^h)$ 
satisfy the inf-sup stability condition 
\eqref{eqn:LBB} and the fluid interface $\Gamma^m$ satisfy the abvoe 
assumptions i)--iii).
Then the numerical method \eqref{eqn:full1}-\eqref{eqn:full3} admits a unique solution.
\end{thm}

\vspace{0.15cm}
\begin{thm}[Unconditional Energy Stability]
\label{th:energylaw}
Assume the electric force is zero. 
Let $\left(\vec u^{m+1}, p^{m+1}, \vec X^{m+1}, \kappa^{m+1}\right)$ 
be the solution to the numerical scheme 
\eqref{eqn:full1}-\eqref{eqn:full3}. 
Then the following stability bound holds
\begin{align}
\tilde{W}(\rho^m,\vec u^{m+1},\vec{X}^{m+1})
+\frac{2\tau}{Re}\norm{\sqrt{\mu^m}D(\vec u^{m+1})}_0^2 
+ \frac{\tau}{Re\cdot l_s}\Bigl(\beta^m u_s^{m+1},~u_s^{m+1}\Bigr)
_{\Gamma^m_1\cup\Gamma_2^m} \nonumber\\
\quad + \frac{\beta^*}{Re\cdot\tau}
\left[\left(x^{m+1}_r-x^m_r\right)^2
+\left(x^{m+1}_l-x^m_l\right)^2\right]\leq
\tilde{W}(\rho^{m-1}, \vec u^m, \vec{X}^m),
\label{eqn:energybounds}
\end{align}
where $\tilde{W}(\rho,\vec u, \vec{X})=\frac{1}{2}(\rho\vec u, \vec u) 
-\frac{\cos\theta_Y}{We}|\Gamma_1| + \frac{1}{We}|\Gamma|$ is the total energy of the system. Moreover, for $m\geq 1$, we have
\begin{align}
\tilde{W}(\rho^{m-1}, \vec u^{m}, \vec{X}^{m})
+\frac{2\tau}{Re}\sum_{k=0}^{m-1}\norm{\sqrt{\mu^k}D(\vec u^{k+1})}_0^2 
+\frac{\tau}{Re\cdot l_s}\sum_{k=0}^{m-1}\left(\beta^k u_s^{k+1}, u_s^{k+1}\right)
_{\Gamma^k_1\cup\Gamma_2^k}\nonumber\\
+ \frac{\beta^*}{Re\cdot\tau}\sum_{k=0}^{m-1}
\left[\left(x^{k+1}_r-x^k_r\right)^2+\left(x^{k+1}_l-x^k_l\right)^2\right]\leq
\tilde{W}(\rho^{0}, \vec u^0, \vec{X}^0).
\label{eqn:discreteenergylaw}
\end{align}
\end{thm}

The three summation terms in \eqref{eqn:discreteenergylaw} 
correspond to the energy dissipation due to the viscous stress in the bulk, the friction force on the wall and the contact line friction, respectively.  

The above theorems are extensions of the previous work by Barrett et al. \cite{Barrett15stable} to problems with moving contact lines. 
In another related work \cite{Zhao19}, similar results were obtained for moving contact lines but on fitted meshes. 
There the energy stability only holds locally in time
due to the required interpolation of the velocity and density between the fitted meshes at each time step. Here the global energy stability of the discrete system is established on the unfitted mesh.
The proof of the theorems is similar to that in Ref. \cite{Zhao19} and is provided in the Appendix \ref{app:th1} and \ref{app:th2}.

The discrete scheme \eqref{eqn:full2}-\eqref{eqn:full3} introduces 
an implicit tangential velocity for the mesh points along the fluid interface 
such that they tend to be uniformly distributed according to the arc length 
in long time \cite{Barrett07, Zhao20}. In our numerical experiments presented below,
no re-meshing for the fluid interface is needed during the simulation.

\section{Numerical results} \label{sec:num}

In this section, we present some numerical results for EWoD obtained using the proposed numerical method.
We first test the accuracy and convergence 
of the boundary integral method for the computation of the electric force on a given fluid interface in section \ref{subsec:bim}.
Then we test the convergence of the numerical method \eqref{eqn:full1}-\eqref{eqn:full3} using the example of a spreading droplet in section \ref{subsec:int}.
Subsequently, we examine the effect of the different parameters in the model on the equilibrium interface profiles in section \ref{subsec:equi}.
Finally, in section \ref{subsec:app} the numerical method is applied to the spreading and migration dynamics of a droplet on various substrates.

Unless otherwise stated, we will choose $\rho_1=1$, $\rho_2=0.1$, $\mu_1=1$, 
$\mu_2=0.1$, $\beta_1=1$, $\beta_2=0.1$, $\beta^*=0.1$, 
$Re=10$, $Ca=0.1$,  $l_s=0.1$ and the initial velocity $\vec u_0=\vec 0$  in the numerical examples. 
The computational domain occupied by the fluids is $\Omega=[-1,1]\times[0,1]$.

\subsection{Convergence test for the electric force}  \label{subsec:bim}

\begin{figure}[!t]
\centering
\includegraphics[width=0.9\textwidth]{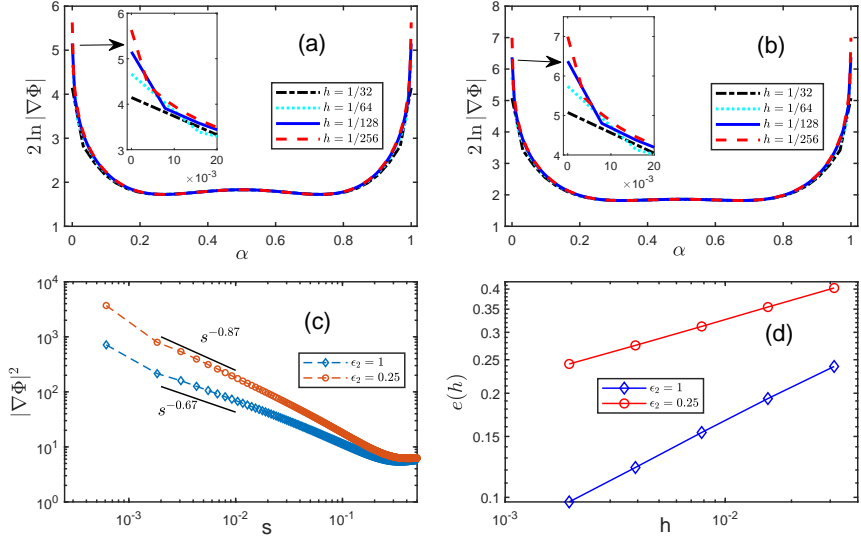}
\caption{{\it Upper panel:} the electric force along the interface (a semi-circle)
computed using different mesh sizes for $\epsilon_2=\epsilon_3=1$ (left) 
and $\epsilon_2=0.25$, $\epsilon_3=1$ (right).
{\it Lower left:} the log-log plot of the electric force versus the arc length computed using $h= 1/2^{10}$.
{\it Lower right: } the relative errors defined in \eqref{eqn:errorphi} versus the mesh sizes $h$. }
\label{fig:density}
\end{figure}

The electrostatic potential $\Phi$ satisfies the Maxwell's equations in the domain $\Omega_2\cup\Omega_3$. This is outside the domain $\Omega_1$, which has a wedge-like geometry with an open angle $\theta$ at the contact points $x_{l,r}$. 
The open angle is equal to the contact angle.
In this geometry, the electric force $|\nabla\Phi|^2 $ in the vicinity 
of the contact point behaves as \cite{Buehrle2003}
\begin{equation}
\label{eqn:asymdensity}
|\nabla\Phi|^2\sim O((\Delta s)^{2(\nu-1)}),\qquad {\rm as}\; \Delta s\to 0^{+},
\end{equation}
where $\Delta s$ is the distance to the contact point, 
and $\nu\in(\frac{1}{2},1)$ is related to the contact angle and satisfies 
\begin{equation}
\label{eqn:asym}
\epsilon_3\tan\left[\nu(\pi-\theta)\right] + 
\epsilon_2 \tan\left(\nu\pi\right) = 0.
\end{equation}
In particular, we have $\nu = \frac{\pi}{2\pi-\theta}$ when $\epsilon_2=\epsilon_3=1$.

To assess the performance of the boundary integral method for the current problem, we compute the electric force along a given interface. The interface is the semi-circle of radius $r=0.4$ centered at $(0,0)$. The thickness of the dielectric substrate is $d=0.2$. The numerical results are shown in 
Fig.~\ref{fig:density}. In the two upper panels, we plot the electric force along the interface. The different curves are obtained using different mesh sizes for
$\epsilon_2=\epsilon_3=1$ (left) and $\epsilon_2=0.25,\ \epsilon_3=1$ (right).
It is evident that the electric force exhibits a singular behavior at the contact points: the force at the contact point (more precisely, the force 
at half grid point away from the contact point) keeps increasing 
as the mesh is refined. 

To further examine the singular structure, we depict the
log-log plot of the electric force against the arc-length in
the lower-left panel of the figure.
 It can be seen that the electric force behaves as $|\nabla\Phi|^2 \sim O(s^p)$ as the contact line is approached.
This is in good agreement with the 
the theoretical prediction of Eqs. \eqref{eqn:asymdensity}-\eqref{eqn:asym}, 
which is shown as the straight lines with slope 
$2(\nu-1)\approx -0.667$ for $\epsilon_2=1$ and 
$2(\nu-1)\approx-0.8718$ for $\epsilon_2=0.25$. 
These lines fit the numerical results very well.

To examine the convergence of the numerical solution as the mesh is refined,
we compute the relative error defined as
\begin{equation}
e(h) = \frac{\int_{\mathbb{I}} \left| |\nabla\Phi^{h}|^2 - |\nabla\Phi^{\frac{h}{2}}|^2\right|
{\rm d}\alpha}{\int_{\mathbb{I}}|\nabla\Phi^\frac{h}{2}|^2 {\rm d}{\alpha}},
\label{eqn:errorphi}
\end{equation}
where $\nabla \Phi^h$ denotes the numerical solution obtained with mesh size $h$.
The error for different mesh sizes is shown in the lower-right panel of Fig. \ref{fig:density}. As expected, the error decreases as the mesh is refined.
In this simulation, we used the uniform mesh along the interface. 
In practice, one may use local mesh refinement near the contact point to obtain more accurate solutions.

\subsection{Convergence test for contact line dynamics}  \label{subsec:int}

\begin{figure}[!t]
\centering
\includegraphics[width=0.95\textwidth]{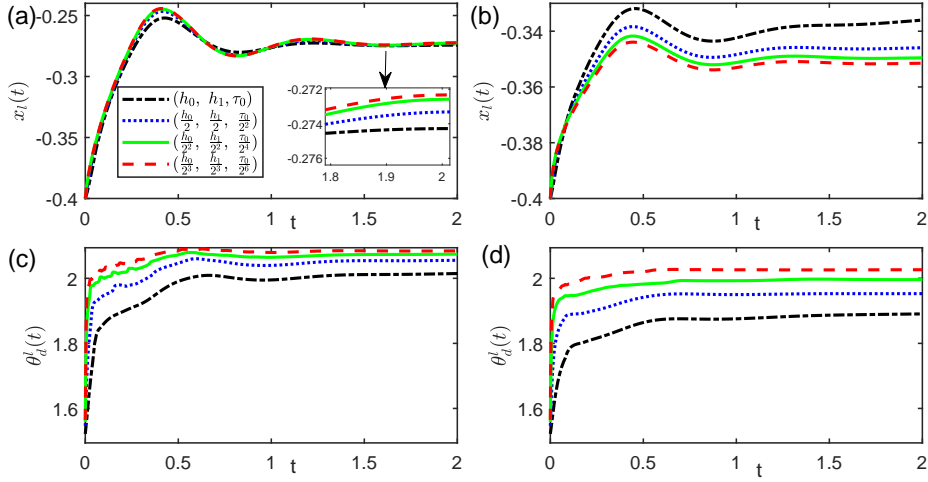}
\caption{The evolution of the (left) contact point (upper panels) and
the contact angle (lower panels) computed using different mesh sizes and time steps.
In the coarse mesh, the mesh size is $h=h_0=1/32$ for the interface and 
$h_\Omega = h_1=1/8$ in the bulk, and the time step is $\tau_0= 0.01$.
Left panels: $\eta=0$; Right panels: $\eta=0.1$.
}
\label{fig:convergence}
\end{figure}

We assess the performance of the numerical method \eqref{eqn:full1}-\eqref{eqn:full3} for the contact line dynamics by carrying out simulations under different mesh sizes.
We consider the spreading dynamics of a droplet. 
Initially the fluid interface is given by a semi-circle with center $(0,0)$ and radius $r=0.4$. The equilibrium contact angle of the droplet is $\theta_Y=2\pi/3$. 
The thickness of the dielectric substrate is $d=0.2$, and the permittivities are 
$\epsilon_2=1$ and $\epsilon_3=1$.

We use uniform mesh (see Fig. \ref{fig:mesh}),
and denote the mesh size in the bulk by $h_\Omega=1/J_{\Omega}$ 
and  the mesh size on the reference domain of the interface $\Gamma$ by $h=1/J_\Gamma$. 
In the boundary integral method, all the boundaries and interfaces are discretized into line segments; for example, the fluid-solid interface $\Gamma_1\cup\Gamma_2$ is also discretized into $J_\Gamma$ line segments.
Different values of $J_\Omega$ and $J_\Gamma$ will be used in the mesh refinement for the convergence test.
For simplicity, we shall fix the discretization of the outer boundaries.

The evolution of the (left) contact point is shown in Fig.~\ref{fig:convergence} 
for $\eta=0$ (upper-left panel) and $\eta=0.1$ (upper-right panel), respectively. In both cases, we can observe the convergence of the contact line dynamics as the mesh and the time step are refined. 
In the lower panels of the same figure, we plot the time history of the contact angle. Similar convergence can also be observed.

\begin{table}[tph]
\caption{Relative change of the droplet area at $t=4$. 
$h,\,h_{_{\Omega}}$ are the mesh size in the discretization of the interface 
and $\Omega$, respectively, and $\tau$ is the time step. 
In the coarse mesh, $h=h_0=1/32$, $h_\Omega=h_1 = 1/8$ 
and $\tau_0=0.01$.}\label{tb:volume}
\begin{center}
\begin{tabular}{@{\extracolsep{\fill}}|c|cc|cc|cc|}\hline
\multirow{2}{*}{$(h,\,h_{_{\Omega}}, \,\tau)$} & \multicolumn{2}{c|}{$\eta=0$} 
&\multicolumn{2}{c|}{$\eta=0.1$} &\multicolumn{2}{c|}{$\eta=0.2$}\\ \cline{2-7}
&$|\Delta A |(t=4) $ & order &$|\Delta A|(t=4)$ 
& order &$|\Delta A|(t=4) $ & order  \\ \hline
$(h_0,\, h_1,\,\tau_0)$ & 2.16E-3 & - &3.38E-4 &-& 3.11E-4 &- \\ \hline
$(\frac{h_0}{2},\,\frac{h_1}{2},\, \frac{\tau_0}{2^2})$ & 6.28E-4 & 1.78 &7.51E-5 &2.17& 1.50E-4 &1.05 
\\ \hline
$(\frac{h_0}{2^2},\, \frac{h_1}{2^2},\,\frac{\tau_0}{2^4})$ & 1.70E-4 & 1.88 &1.77E-5 &2.09& 4.82E-5 &1.64 \\ \hline
$(\frac{h_0}{2^3},\, \frac{h_1}{2^3},\,\frac{\tau_0}{2^6})$ & 4.33E-5 & 1.97 &4.14E-6 &2.10 &1.36E-5 &1.82\\\hline
 \end{tabular}
\end{center}
 \end{table}

 \begin{table}[tph]
\caption{Convergence of the contact angle to the equilibrium angle 
at $t=4$. $h,\,h_{_{\Omega}}$ are the mesh size in the discretization 
of the interface and $\Omega$, respectively, and $\tau$ is the time step. 
In the coarse mesh,  $h=h_0=1/32$, $h_\Omega=h_1 = 1/8$, $\tau_0=0.01$.}
\label{tb:angle}
\begin{center}
 \begin{tabular}{@{\extracolsep{\fill}}|c|cc|cc|cc|}\hline
\multirow{2}{*}{$(h,\,h_{_{\Omega}}, \,\tau)$} & \multicolumn{2}{c|}{$\eta=0$} &\multicolumn{2}{c|}{$\eta=0.1$} &\multicolumn{2}{c|}{$\eta=0.2$}\\ \cline{2-7}
&$|\Delta\theta|(t=4) $ & order &$|\Delta\theta|(t=4)$ 
& order &$|\Delta\theta|(t=4) $ & order  \\ \hline
$(h_0,\, h_1,\,\tau_0)$ & 7.12E-2 & - &2.00E-1 &-& 3.79E-1 &- \\ \hline
$(\frac{h_0}{2},\,\frac{h_1}{2},\, \frac{\tau_0}{2^2})$ & 3.54E-2 & 1.01 &1.40E-1 &0.51& 3.11E-1 &0.29 
\\ \hline
$(\frac{h_0}{2^2},\, \frac{h_1}{2^2},\,\frac{\tau_0}{2^4})$ & 1.77E-2 & 1.00 &9.74E-2 &0.52& 2.27E-1 &0.45 \\ \hline
$(\frac{h_0}{2^3},\, \frac{h_1}{2^3},\,\frac{\tau_0}{2^6})$ & 8.90E-3 & 0.99 &6.61E-2 &0.55 &1.53E-1  &0.57\\ \hline
 \end{tabular}
\end{center}
 \end{table}

Next we examine the conservation of area for the droplet. 
The finite element space $\mathbb{P}^h$ for the pressure given in \eqref{eqn:spaceUP} 
contains piecewise constant functions, which ensures 
the local mass conservation particularly over each element. Besides, by choosing $\zeta = \chi_{_{\Omega_1(t)}}$ in \eqref{eqn:weak2a} and $\psi = 1$ in \eqref{eqn:weak2}, we can establish the mass conservation law within the weak form as
\begin{align}
&\frac{{\rm d}}{{\rm d}t}|\Omega_1(t)| = \Bigl(\vec n\cdot\partial_t\vec X,~1\Bigr)_{\Gamma(t)}
\nonumber\\
&\quad=\Bigl(\vec u\cdot\vec n,~1\Bigr)_{\Gamma(t)}=\int_{\Omega_1(t)}\nabla\cdot\vec u\;\dL^2= \Bigl(\nabla\cdot\vec u,~\chi_{_{\Omega_1(t)}}\Bigr)=0.
\end{align}
Therefore in this and the following simulations, 
we further enrich the pressure space $\mathbb{P}^h$ with the basis function $\chi_{_{\Omega_1^m}}$, the characteristic function over the domain occupied by the droplet. This helps preserving the area of the droplet \cite{Barrett15stable}.
In addition, the pressure jump across the interface can be captured with this function. In practical numerical implementations, the new contribution of the single basis function to \eqref{eqn:full1} and \eqref{eqn:full2a} can be written in terms of the integrals over $\Gamma^m$ as
\begin{equation}
\Bigl(\nabla\cdot\boldsymbol{\omega}^h,~\chi_{_{\Omega_1^m}}\Bigr) = \int_{\Omega_1^m}\nabla\cdot\boldsymbol{\omega}^h\dL^2 = \left(\boldsymbol{\omega}^h,~\vec n^m\right)_{\Gamma^m}^h,\qquad \forall\boldsymbol{\omega}^h\in \mathbb{U}^h.
\end{equation}

In  Table.~\ref{tb:volume}, we present the relative area change $|\Delta A|$ of the droplet at $t=4$. At this time the droplet has evolved close to the steady state. Here, $\Delta A$ is defined as
\begin{equation}
\label{eqn:numqu}
\Delta A^h(t_m) = \frac{|\Omega_1^m| -|\Omega_1^0|}{|\Omega_1^0|},
\end{equation}
where $|\Omega_1^m|$ is the area of the droplet at time $t_m$.
From the table, we observe that the droplet area is well-preserved. 
As the mesh is refined, the area change $|\Delta A|$ converges to zero with order close to 2.

After the droplet reaches the steady state, the theoretical value of the contact angle should converge to the equilibrium angle $\theta_Y=2\pi/3$. 
In table \ref{tb:angle}, we present $\Delta \theta$, 
the deviation of the contact angle from this equilibrium angle at $t=4$, obtained with different mesh sizes.
 We observe that in all three cases for the different values of $\eta$, the contact angle indeed converges to the equilibrium angle. 
When $\eta=0$, i.e. in the absence of electric field, 
the convergence order is close to 1; however, the order 
is reduced to about 0.5 for $\eta=0.1,\;0.2$.

The order of convergence for the contact angle can be understood as follows. 
Denote by $\Gamma^h=\vec X^h$ and $\kappa^h$ the numerical solution for the interface and its curvature at the steady state, respectively.
Then from \eqref{eqn:full3}, we have 
\begin{equation}
\label{eqn:ddsy}
\Bigl(\kappa^h\,\vec n^h,~\boldsymbol{g}^h\Bigr)_{\Gamma^h}^h
+ 
\Bigl(\partial_s\vec X^h,~\partial_s\boldsymbol{g}^h\Bigr)_{\Gamma^h}^h
-\cos\theta_Y\left[g_1^h(1) - g_1^h(0)\right]=0,
\quad\forall\, \b{g}^h\in \mathbb{X}^h.
\end{equation}
By choosing $\b{g}^h=(g_1^h, ~g_2^h)^T:=(\phi_0(\alpha),~0)^T$ in \eqref{eqn:ddsy}, 
where $\phi_0(\alpha)$ is the piecewise linear function taking the value
1 at $\alpha_0=0$ and 0 at all other nodes, we obtain 
\begin{equation}
\frac{1}{2}\kappa^h(0)n_{1, 1}^h\left|\vec X^h(\alpha_1)-\vec X^h(\alpha_0)\right|
-\cos\theta^h +\cos\theta_Y =0,
\end{equation}
where $n_{1,1}^h$ is the first component of $\vec{n}_1^h$, and $\theta^h$
is the contact angle of $\vec{X}^h$.
This yields
\begin{equation}
\left|\cos\theta^h- \cos\theta_Y\right|
= \frac{1}{2}\left|\kappa^h(0)\sin\theta^h\right|\Delta s,
\end{equation}
where $\Delta s = \left|\vec X^h(\alpha_1)-\vec X^h(\alpha_0)\right|$. 
On the other hand, the interface condition implies that
the curvature of the fluid interface has the same singular structure 
as the electric field at the contact point. 
Therefore, we have $\kappa^h(0)\sim O\left((\Delta s)^{2(\nu-1)}\right)$, 
and consequently
\begin{equation}
\left|\theta^h - \theta_Y\right| \leq C_1(\Delta s)^{2\nu-1}
\leq C_2 h^{2\nu-1}, 
\end{equation}
where $C_1$ and $C_2$ are constants independent of the mesh size $h$. 
When $\eta=0$, we have $\nu=1$ and the curvature is constant 
along the interface at the steady state, therefore the convergence order of the contact angle is 1. On the other hand, in the presence of the electric field, we have $1/2< \nu< 1$, and the convergence order is lowered. 
In the current example,  we have $\nu=3/4$ from  Eq. \eqref{eqn:asym}, 
thus $\left|\theta^h - \theta_Y\right| \sim O(h^{1/2})$, 
which is consistent with the numerical results.

\subsection{Equilibrium interface profiles} \label{subsec:equi}

\begin{figure}[!t]
\centering
\includegraphics[width = 0.90\textwidth]{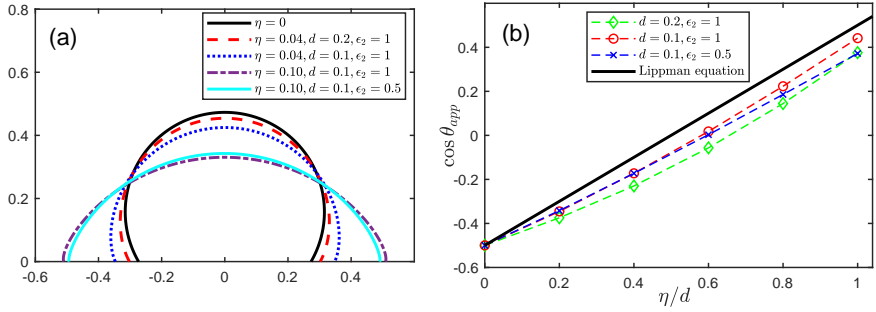}
\caption{{\it Left panel}: Equilibrium profiles of the interface.
{\it Right panel}: The value of $\cos\theta_{app}$ versus $\eta/d$ (discrete points with dash lines) and the prediction of Lippmann equation (straight line).}
\label{fig:ew1}
\end{figure}

In this example, we investigate the influence of the various model parameters, such as 
$\eta$, $\epsilon_2$ and $d$, on the equilibrium profile of the fluid interface.
The fluid interface of the droplet is initially given by a semi-circle with centre $(0,0)$ and radius $r=0.4$. The equilibrium contact angle of the droplet
is $\theta_Y=2\pi/3$. 
The fluid domain $\Omega = [-1, 1]\times [0, 1]$ is discretized into $4848$ triangles
with $2534$ vertices, and the fluid interface is discretized into $J_{\Gamma}=512$ elements. The time step is $\tau=2\times 10^{-4}$.

The numerical results are shown in Fig.~\ref{fig:ew1}.
The left panel shows the equilibrium profiles of the interface for different values of 
$\eta$, $\epsilon_2$ and $d$. 
Comparing with the interface profile when $\eta=0$ 
(i.e. in the absence of electric field), we see that the electric force flattens the interface and make the droplet spread.

A more quantitative assessment of the effect of the electric force is shown 
in the right panel, where we plot $\cos\theta_{app}$ against $\eta/d$ for different values of $\epsilon_2$ and $d$.
The apparent contact angle $\theta_{app}$ is computed by fitting the interface by a circular arc using the apex of the interface and the given area of the droplet. 
From the numerical results, we observe that the ratio $\eta/d$ plays the dominant role here; more specifically, $\cos\theta_{app}$ increases linearly with
$\eta/d$ with the slope close to 1. In contrast, 
the permittivity of the fluid outside the droplet, $\epsilon_2$, has little effect on the interface profile. 
This is in good agreement with the Lippmann equation
\eqref{eqn:Liapp}. In terms of the dimensionless parameters, 
Eq. \eqref{eqn:Liapp} reads
\begin{equation} \label{eq:Lipp1}
\cos\theta_{B}=\cos\theta_Y + \frac{\eta}{d}.
\end{equation}
The contact angle $\theta_B$ computed using this equation is also shown 
in the figure, and good agreement with the numerical results can be observed.
The discrepancy might be due to the finite system size, the effect of the boundary conditions, or the finite value of $d$ and $\eta$. 
After all, the Lippmann equation
is an asymptotic result which is derived in the limit of $d\rightarrow 0$ and $\eta\rightarrow 0$ \cite{Cui2019}.

\begin{figure}[!t]
\centering
\includegraphics[width = 0.90\textwidth]{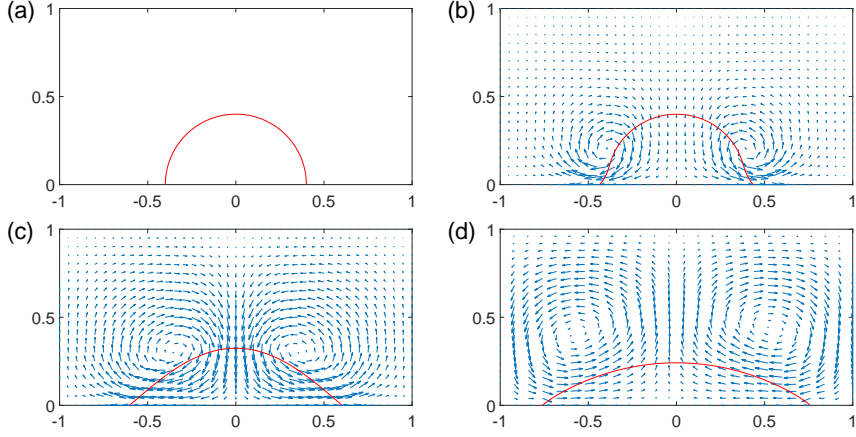}
\caption{Snapshots of the interface and the velocity field
on a hydrophilic dielectric substrate with $\theta_Y=\pi/3$.
Parameters are $\eta=0.125$, $\epsilon_2=\epsilon_3=1$ and $d=0.2$. 
(a) $t=0$; (b) $t=0.06$, $\max_{\vec x\in\Omega}|\vec u| = 1.027$ (c) $t=0.2$, $\max_{\vec x\in\Omega}|\vec u|=0.740$; (d) $t=1.5$, $\max_{\vec x\in\Omega}|\vec u| = 0.004$.
 }
\label{fig:wetting}
\end{figure}
\begin{figure}[!tph]
\centering
\includegraphics[width = 0.90\textwidth]{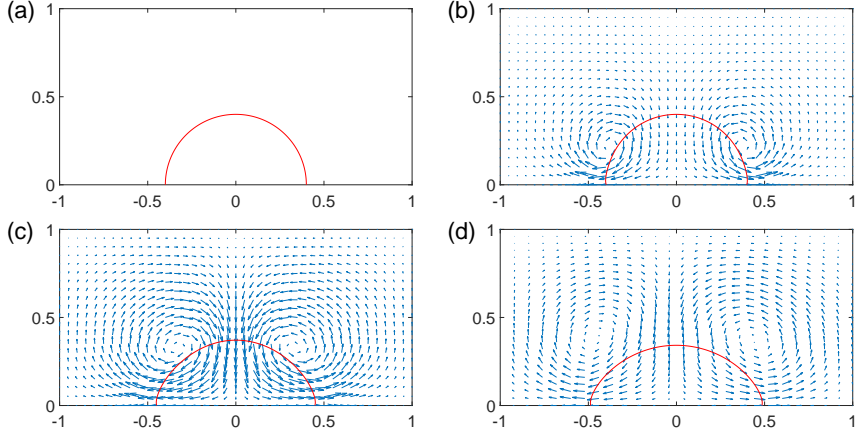}
\caption{Snapshots of the interface and the velocity field on
a hydrophobic dielectric substrate with $\theta_Y=2\pi/3$.
Parameters are $\eta=0.2$, $\epsilon_2=\epsilon_3=1$ and $d=0.2$. 
(a) $t=0$; (b) $t=0.06$, $\max_{\vec x\in\Omega}|\vec u|=0.282$; 
(c) $t=0.2$, $\max_{\vec x\in\Omega}|\vec u|=0.284$; (d) $t=1.5$, $\max_{\vec x\in\Omega}|\vec u|=0.003$.
}
\label{fig:dewetting}
\end{figure}

\begin{figure}[!tph]
\centering
\includegraphics[width = 0.90\textwidth]{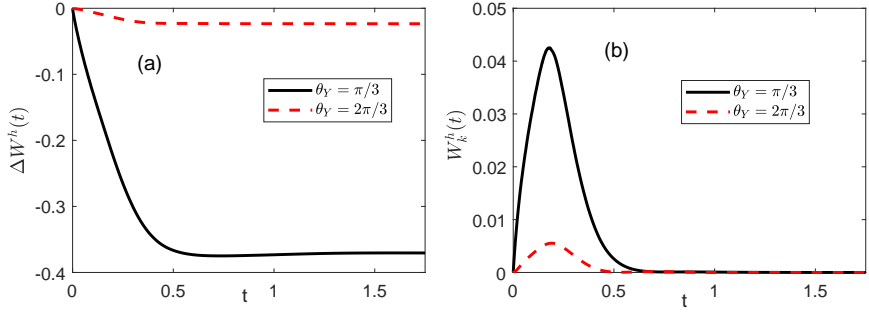}
\caption{The energy loss $\Delta W(t) = W (t) - W(0)$ (left panel) and the kinetic energy 
$W_k(t) = \frac{1}{2}\int_{\Omega}\rho|\vec u|^2 d\mathcal{L}^2$ (right panel) as functions of time. Here the discrete energy $W(t)$ is computed using the dimensionless from of \eqref{eqn:alterdimensionEnergy}.} 
\label{fig:evolution}
\end{figure}

\subsection{Applications}  \label{subsec:app}

We investigate the detailed dynamics of a droplet on dielectric substrates driven by the electric force.
We consider a hydrophilic case with the contact angle $\theta_Y=\pi/3$ and a hydrophobic case with the contact angle $\theta_Y=2\pi/3$. 
The initial configuration of the system 
and the discretizations of the computational domain and the interface are the same as those in the previous example.
Several snapshots of the interface profile are shown in Fig.~\ref{fig:wetting} 
for the hydrophilic case and in Fig.~\ref{fig:dewetting} for the hydrophobic case. 
Also shown in the figures are the respective velocity field. 
We observe that vortices are generated near the contact points, driving the droplet to spread on the substrate.

In Fig.~\ref{fig:evolution}, we plot the loss of the total energy 
$\Delta W: =W(t)-W(0)$ (left panel) 
and the kinetic energy of the fluids 
$W_k(t) = \int_{\Omega}\frac{1}{2}\rho |\vec{u}|^2 d\mathcal{L}^2$  (right panel)
as functions of time. 
The total energy, as given in Eq. \eqref{eqn:dimenenergy}, consists of the kinetic energy, the interfacial energies and the electrostatic energy.
We observe that the total energy of the discrete system decays in time, a desired property of the numerical method.

\begin{figure}[tph]
\centering
\includegraphics[width = 0.95\textwidth]{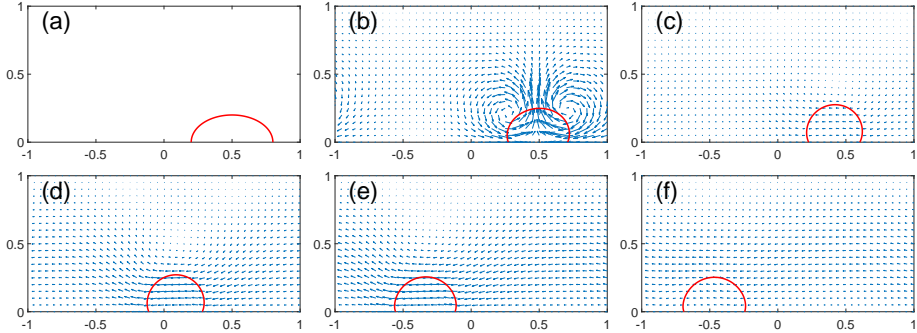}
\caption{Snapshots of the droplet migrating on a dielectric substrate.  
The electrostatic potential on the lower boundary of the substrate is prescribed 
as $\Phi|_{y=-d} = \frac{3}{8} (x +1)$. Other parameters are 
$\eta = 0.2$, $\epsilon_2=\epsilon_3=1$, $d=0.2$, and $\theta_Y=2\pi/3$. (a) $t=0$; (b) $t=0.1$; $\max_{x\in\Omega}|\vec u| = 0.99$; (c) $t=0.5$; $\max_{\vec x\in\Omega}|\vec u| = 0.34$; (d) $t=1.2$, $\max_{\vec x\in\Omega}|\vec u| = 0.56$; (e)  $t=1.8$, $\max_{x\in\Omega}|\vec u|=0.90$; (f)$t=2.3$, $\max_{\vec x\in\Omega}|\vec u|=0.41$.}
\label{fig:migration}
\end{figure}

In the last example, we simulate the migration of a droplet on a substrate with non-uniform electrostatic potential prescribed on the boundary. 
The non-uniform potential mimics an array of electrodes placed below the substrate that is used to transport the droplet in experiments. 
The electrostatic potential on the lower boundary of the substrate is given as
$\Phi|_{y=-d} = \frac{3}{8}(x+1)$. 
Other parameters are chosen as $\eta=0.2$, $\epsilon_2=\epsilon_3=1$, $d=0.2$, 
and $\theta_Y=2\pi/3$. 
The initial interface of the droplet is given by a semi-ellipse as $\frac{(x-0.5)^2}{0.3^2} + \frac{y^2}{0.2^2}=1,\;y\geq 0$. 
Numerical results for the interface profile and the velocity field are shown in Fig.~\ref{fig:migration}. 
We observe that the droplet first de-wets the substrate to form a near-circular shape with contact angle close to $\theta_Y=2\pi/3$. 
In this stage, the dynamics is mainly driven by the surface tension. 
Afterwards, the electric force plays dominant role, and it drives the droplet to migrate from the right to the left.

\section{Conclusions}
\label{sec:con}

In this work, we presented a hydrodynamic model for electrowetting on dielectric
based on the earlier work on moving contact lines, and developed an efficient numerical method for the model.
The numerical method combines a semi-implicit parametric finite element method 
for the dynamics of the fluid interface and the finite element method for the Navier-Stokes equations as well as the boundary integral method for the electric field. We proved that the numerical method admits a unique solution.
In the case without the electric field, we showed that the numerical method obeys a similar energy law as the continuum model. 

In the numerical experiments, we assessed the accuracy and convergence 
of the numerical method and investigated the effect of the different
physical parameters on the interface dynamics and its equilibrium profile. 
Numerical results for the equilibrium profile of the interface agree well with the predictions of the Lippmann equation. 

The numerical solution for the electric force 
exhibits a singular structure near the contact point that is consistent with theoretical results. This singularity incurs large curvature of the interface
near the contact point, which deteriorates the convergence order 
of the numerical solution, particularly the convergence of the contact angle to the equilibrium angle.   

In this work, we focused on simulations in two dimensions. 
The numerical method can be readily extended to three-dimensional problems. This will be left to our future work.

\section*{Acknowledgement}
The work was partially supported by Singapore MOE RSB grant, 
Singapore MOE AcRF grants (No. R-146-000-267-114, No. R-146-000-285-114) 
and NSFC grant (No. 11871365).

\appendix
\section{Energy law for the continuum model}
\label{ap:energylaw}
The total energy for the EWoD model \eqref{eqn:fluidynamic}-\eqref{eqn:phi4} in its original dimension form reads
\begin{equation}
W(t)=\sum_{i=1,2}\int_{\Omega_i}\frac{1}{2}\rho_i|\vec u|^2\dL^2- \gamma\cos\theta_Y|\Gamma_1(t)| +\gamma|\Gamma(t)| - \sum_{i=2,3}\int_{\Omega_i}\frac{1}{2}\epsilon_i|\nabla\Phi|^2\;\dL^2.
\label{eqn:dimensionEnergy}
\end{equation}
Integrating by parts and using the electrostatic potential equation in \eqref{eqn:phi1} as well as the boundary conditions for $\Phi$, we can transform the electrical energy in \eqref{eqn:dimensionEnergy} into expressions only involving line integrals over $\Gamma$ and $\Gamma_1$. This gives the following alternative form of the total energy
\begin{align}
&W(t)=\sum_{i=1,2}\int_{\Omega_i}\frac{1}{2}\rho_i|\vec u|^2\dL^2- \gamma\cos\theta_Y|\Gamma_1(t)| +\gamma|\Gamma(t)| \nonumber\\
&\quad\qquad\qquad\qquad + \frac{\phi}{2}\left(\int_{\Gamma}\epsilon_2\,(\vec n\cdot\nabla\Phi)\;\ds + \int_{\Gamma_1}\epsilon_3\,(\vec n_w\cdot\nabla\Phi)\;\ds\right),
\label{eqn:alterdimensionEnergy}
\end{align}
which can be used to compute the discrete electrical energy in view of the boundary integral method in \eqref{eqn:disbem}. The dynamic system obeys the energy dissipation law 
\begin{equation}
\frac{{\rm d}}{{\rm d}t}W(t) = -\sum_{i=1,2}\int_{\Omega_i}2\mu_i\norm{D(\vec u)}_F^2\dL^2 - \sum_{i=1,2}\int_{\Gamma_i}\beta_i|u_s|^2\ds - \beta^*(\dot{x}_l^2 + \dot{x}_r^2)\leq 0.
\label{eqn:energylawapp}
\end{equation}
We show the proof of \eqref{eqn:energylawapp}. The dissipation of the fluid kinetic energy is 
\begin{align}
&\frac{{\rm d}}{{\rm d}t}\sum_{i=1,2}\int_{\Omega_i(t)}\frac{1}{2}\rho_i|\vec u|^2\;\dL^2=\sum_{i=1,2}\int_{\Omega_i}\rho_i\vec u\cdot(\partial_t\vec u + \vec u\cdot\nabla\vec u)\;\dL^2 \nonumber\\
&\quad= \sum_{i=1,2}\int_{\Omega_i}\vec u\cdot(-\nabla p + \nabla\cdot\tau_d)\;\dL^2\nonumber\\
&\quad= -\sum_{i=1,2}\int_{\Omega_i}\nabla\vec u:\tau_d\;\dL^2 -\sum_{i=1,2}\int_{\Gamma_i}\vec u\cdot\tau_d\cdot\vec n_w\;\ds+\int_{\Gamma}\vec u\cdot[p\mathbf{I}_2-\tau_d]_1^2\cdot\vec n\;\ds \nonumber\\ 
&\quad=-\sum_{i=1,2}\int_{\Omega_i}2\mu_i\norm{D(\vec u)}_F^2\;\dL^2  - \sum_{i=1,2}\int_{\Gamma_i}\beta_i|u_s|^2\;\ds \nonumber\\
&\qquad\qquad\qquad\qquad+\;\int_{\Gamma}(\gamma\kappa+ \frac{\epsilon_2}{2}|\nabla\Phi|^2)(\vec u\cdot\vec n)\;\ds,\label{eq:dissipation1}
\end{align}
where we have used the divergence free condition in \eqref{fluidynamic2}, the interface conditions in \eqref{eqn:bd1a}-\eqref{eqn:bd1b}, the boundary condition \eqref{eqn:bdp2}, the no-slip boundary condition on the upper wall $\Gamma_4$ and the periodic boundary conditions along $\Gamma_3$.

The time derivative of the interfacial energies is
\begin{align}
&\frac{{\rm d}}{{\rm d}t}\Bigl(-\gamma\cos\theta_Y|\Gamma_1(t)| + \gamma|\Gamma(t)|\Bigr)\nonumber\\
&\quad = -\gamma\cos\theta_Y(\dot{x_r} - \dot{x_l}) - \int_\Gamma \gamma\kappa\, v_n\;\ds + \gamma\left(\dot{x_r}\cos\theta_d^r - \dot{x_l}\cos\theta_d^l\right)\nonumber\\
&\quad = -\int_\Gamma\gamma\kappa\,(\vec u\cdot\vec n)\;\ds - \beta^*\left(\dot{x_r}^2 + \dot{x_l}^2\right),
\label{eq:dissipation2}
\end{align}
where we have used \eqref{eqn:bd1c} and \eqref{eqn:bdp3}.

Denote by $\hat{\vec n}$ the outward unit normal vector on the boundary of $\Omega_2$ and $\Omega_3$. Differentiating the electrical energy yields
\begin{align}
&\frac{{\rm d}}{{\rm d}t}\left(-\frac{1}{2}\sum_{i=2,3}\int_{\Omega_i}\epsilon_i|\nabla\Phi|^2\;\dL^2\right)\nonumber\\
&\quad = -\frac{1}{2}\sum_{i=2,3}\int_{\Omega_i}2\epsilon_i\nabla\Phi\cdot\nabla(\partial_t\Phi)\;\dL^2 - \frac{1}{2}\sum_{i=2,3}\int_{\partial\Omega_i}\epsilon_i|\nabla\Phi|^2(\vec u\cdot\hat{\vec n})\,\ds\nonumber\\
&\quad = -\frac{1}{2}\sum_{i=2,3}\int_{\Omega_i}2\epsilon_i\nabla\cdot(\partial_t\Phi\nabla\Phi)\;\dL^2 + \frac{1}{2}\int_{\Gamma}\epsilon_2|\nabla\Phi|^2(\vec u\cdot\vec n)\;\ds\nonumber\\
&\quad = -\sum_{i=2,3}\int_{\partial\Omega_i}\epsilon_i\partial_t\Phi\,(\nabla\Phi\cdot\hat{\vec n})\,\ds+ \frac{1}{2}\int_{\Gamma}\epsilon_2|\nabla\Phi|^2(\vec u\cdot\vec n)\;\ds\nonumber\\
&\quad = -\frac{1}{2}\int_{\Gamma}\epsilon_2|\nabla\Phi|^2\,(\vec u\cdot\vec n)\;\ds,
\label{eq:dissipation3}
\end{align}
where the first equality results from the Reynolds transport theorem, the second equality is obtained from \eqref{eqn:phi1}, the third equality comes from the divergence theorem, and for the last equality, we have used the boundary conditions in \eqref{eqn:phi2}-\eqref{eqn:phi4}, the periodic boundary condition along $\Gamma_3\cup\Gamma_6$ as well as the fact that 
\begin{equation}
\partial_t\Phi +\vec u\cdot\nabla\Phi = 0,\qquad{\rm on}\;\Gamma\cup\Gamma_1\cup\Gamma_4\cup\Gamma_5.
\end{equation}
Combining Eqs.~\eqref{eq:dissipation1}-\eqref{eq:dissipation3}, we obtain the energy dissipation law in \eqref{eqn:energylawapp}. In terms of the dimensionless variables defined in section \ref{sec:dimensionless}, the energy law in \eqref{eqn:energylawapp} gives Eq.~\eqref{eqn:dimensionenergydissipation2d} (scaled by $\rho_1U^2L^2$).

\section{Proof of Theorem \ref{th:wellposed}}
\label{app:th1}
It suffices to show that the corresponding homogeneous system has only zero solution. By noting that electric force $\epsilon_2\eta|\nabla\Phi|^2$ in \eqref{eqn:full1} is explicitly evaluated on $\Gamma^m$, thus we can consider solving the following homogeneous system for
$\big(\vec u^h,~p^h,~\vec X^h,~\kappa^h\big)\in
\big(\mathbb{U}^h,~\mathbb{P}^h,~\mathbb{X}^h,~\mathbb{K}^h\big)$ such that
\begin{subequations}
\begin{align}
\label{eqn:homoge1}
&\frac{1}{2}\Bigl[\Bigl(\frac{(\rho^m+\rho^{m-1})\vec u^h}{\tau},
~\boldsymbol{\omega}^h\Bigr) +\Bigl(\rho^m(\vec u^m\cdot\nabla)
\vec u^h,~\boldsymbol{\omega}^h\Bigr)\nonumber\\&\quad-\;\Bigl(\rho^m(\vec u^m\cdot\nabla)
\boldsymbol{\omega}^h,~\vec u^h\Bigr)\Bigr]-\Bigl(p^h,~\nabla\cdot\boldsymbol{\omega}^h\Bigr)\nonumber\\
&\quad
+\frac{2}{Re}\,\Bigl(\mu^m D(\vec u^h),~D(\boldsymbol{\omega}^h)\Bigr)
-\;\frac{1}{We}\Bigl(\kappa^h\,\vec n^m,~\boldsymbol{\omega}^h\Bigr)_{\Gamma^m}\nonumber\\ 
&\quad +\;\frac{1}{Re\cdot l_s}\Bigr(\beta^m\,u_s^h,~\omega_s^h\Bigr)_
 {\Gamma_1^m\cup\Gamma_2^m}=0,\quad\forall\boldsymbol{\omega}^h\in \mathbb{U}^h,\\[0.5em]
 \label{eqn:homoge2}
&\qquad\qquad\Bigl(\nabla\cdot\vec u^h,~\zeta^h\Bigr)=0,\quad\forall\zeta^h\in \mathbb{P}^h,\\[0.5em]
\label{eqn:homoge3}
&\Bigl(\frac{\vec X^h}{\tau}\cdot\vec n^m,~\psi^h\Bigr)
_{\Gamma^m}^h - \Bigl(\vec u^h\cdot\vec n^m,~\psi^h\Bigr)_{\Gamma^m}=0,
\quad\forall \psi^h\in \mathbb{K}^h,\\[0.5em]
&\Bigl(\kappa^h\,\vec n^m,~\boldsymbol{g}^h\Bigr)_{\Gamma^m}^h
+\Bigl(\partial_s\vec X^h,~\partial_s\boldsymbol{g}^h\Bigr)_{\Gamma^m}^h\nonumber\\
&\qquad\qquad\qquad 
+\frac{\beta^*Ca}{\tau}\Bigl[x_r^h\,g_1^h(1)+x_l^h\,g_1^h(0)\Bigr]=0,
\quad\forall\boldsymbol{g}^h\in \mathbb{X}^h,
\label{eqn:homoge4}
\end{align}
\end{subequations}
where $\vec X^h=(X^h,~Y^h)^T$, $u_s^h = \vec u^h\cdot\vec t_w$, and $x_l^h:=X^h|_{\alpha=0}$ and $x_r^h:=X^h|_{\alpha=1}$.

Taking $\boldsymbol{\omega}^h=\vec u^h$, $\zeta^h = p^h$, 
$\psi^h=\frac{1}{We}\kappa^h$ and $\boldsymbol{g}^h = \frac{1}{We}\vec X^h$, then
combining these equations yields
\begin{align}
&\Bigl(\frac{\rho^m+\rho^{m-1}}{2}\vec u^h,\vec u^h\Bigr) 
+ \frac{2\tau}{Re}\Bigl(\mu^m\,D(\vec u^h),D(\vec u^h)\Bigr) 
+ \frac{\tau}{Re\cdot l_s}\Bigl(\beta^m\,u_s^h,u_s^h\Bigr)
_{\Gamma_1^m\cup\Gamma_2^m} \nonumber\\
& \qquad\qquad
+ \frac{1}{We}\Bigl(\partial_s\vec X^h,~\partial_s\vec X^h
\Bigr)_{\Gamma^m}^h+ \frac{\beta^{*}}{Re\cdot \tau}[(x_r^h)^2+(x_l^h)^2]=0.
\end{align}
By Korn's inequality, we have
\begin{equation}
\norm{\vec u^h}_1\leq C\Bigl[\frac{1}{2}\Bigl((\rho^m+\rho^{m-1})\vec u^h,~\vec u^h\Bigr) 
+ \frac{2\tau}{Re}\Bigl(\mu^m\,D(\vec u^h),~D(\vec u^h)\Bigr)\Bigr ]\leq 0,
\end{equation}
thus we immediately obtain $\vec u^h=\vec 0$. We also have $\vec X^h=\vec 0$ by noting $x_r^h=x_l^h=0$. Substituting $\vec X^h=\vec 0$ into Eq.~\eqref{eqn:homoge4}, we obtain 
\begin{equation}
\label{eqn:kkzero}
\Bigl(\kappa^h\,\vec n^m,~\boldsymbol{g}^h\Bigr)_{\Gamma^m}^h=0,\qquad 
\forall\boldsymbol{g}^h\in \mathbb{X}^h.
\end{equation}
Denote $\vec n_j^m=(n_{j,1}^m,~n_{j,2}^m)^T,\; j = 1,2,\cdots,J_{_\Gamma}$. Choosing $\boldsymbol{g}^h$ in \eqref{eqn:kkzero} as
\begin{equation}
\left. \boldsymbol{g}^h\right|_{\alpha_j} =\left\{
\begin{array}{l}
\left[\vec h_{j+1}^m +\vec h_j^m)\right]^\perp \kappa^h(\alpha_j),
\quad 1\leq j\leq J_{_\Gamma}-1, \vspace{0.15cm}\\
\left(n_{1,1}^{m}\kappa^h(\alpha_j), ~0\right)^T, \quad j = 0,\vspace{0.15cm} \\
\left(n_{J_{_\Gamma},1}^{m}\,\kappa^h(\alpha_j),~0\right)^T,\quad j = J_{_\Gamma},
\end{array} \right.
\end{equation}
and by noting the norm in \eqref{eqn:massnorm}, we obtain
\begin{align}
0=&\,\frac{(\kappa^h(\alpha_0))^2 }{2}|\vec h_1^m|(n_{1,1}^m)^2+ \frac{(\kappa^h(\alpha_{J_{_\Gamma}}))^2}{2}|\vec h_{J_{_\Gamma}}^m|(n_{J_{_\Gamma},1}^m)^2\nonumber\\
&\qquad+\;\frac{1}{2}\sum_{j=1}^{J_{_\Gamma}-1}\left(\kappa^h(\alpha_j)\right)^2 |\vec h_j^m + \vec h_{j+1}^m|^2,
\end{align}
which implies $\kappa^h(\alpha_j)=0,\;\forall 0\leq j\leq J_{_\Gamma}$ from the assmuptions i)--iii).  We then substitute $\vec u^h = \vec 0$ and $\kappa^h=0$ 
into \eqref{eqn:homoge1} and obtain
 \begin{equation}
 \left(p^h,~\nabla\cdot\boldsymbol{\omega}^h\right)=0,
\qquad\forall \boldsymbol{\omega}^h\in \mathbb{U}^h.
 \end{equation}
Using the stability condition in \eqref{eqn:LBB}, we consequently 
obtain $p^h=0$. This shows that the homogeneous linear system 
\eqref{eqn:homoge1} - \eqref{eqn:homoge4} has only the zero solution.  
Thus, the numerical scheme \eqref{eqn:full1}-\eqref{eqn:full3}
admits a unique solution.

\section{Proof of Theorem \ref{th:energylaw}}
\label{app:th2}
Setting $\boldsymbol{\omega}^h=\vec u^{m+1}$, $\zeta^h = p^{m+1}$, 
$\psi^h = \frac{1}{We}\,\kappa^{m+1}$ and 
$\boldsymbol{g}^h = \frac{1}{We\cdot\tau}(\vec X^{m+1}-\vec X^m)$ in Eqs.
\eqref{eqn:full1}-\eqref{eqn:full3}, noting $\eta= 0$ and then combining these equations yield
\begin{align}
&\frac{1}{2\tau }\left[\Bigl(\rho^m\vec u^{m+1}-\rho^{m-1}\vec u^m,
~\vec u^{m+1}\Bigr)+\Bigl(\rho^{m-1}\left(\vec u^{m+1}-\vec u^m\right),
~\vec u^{m+1}\Bigr)\right]\nonumber\\  
&\qquad+\;\frac{2}{Re}\,\Bigl(\mu^m D(\vec u^{m+1}),~D(\vec u^{m+1})\Bigr)+\frac{1}{Re\cdot l_s}\left(\beta^m\,u_s^{m+1},~u_s^{m+1}\right)
_{\Gamma_1^m\cup\Gamma_2^m}\nonumber\\
&\qquad +\;\frac{1}{We\cdot\tau}\Bigl(\partial_s\vec X^{m+1},
~\partial_s(\vec X^{m+1}-\vec X^m)\Bigr)_{\Gamma^m}^h\nonumber\\
&\qquad-\;\frac{\cos\theta_Y}{We\cdot\tau}\left[(x_r^{m+1}-x_l^{m+1}) -(x_r^m-x_l^m)\right]\nonumber\\
&\qquad+\;\frac{\beta^*}{Re} \left[\left(\frac{x_r^{m+1}-x_r^m}{\tau}\right)^2
+\left(\frac{x_l^{m+1}-x_l^m}{\tau}\right)^2\right]=0.
\label{eqn:energybound1}
\end{align}
It is easy to see that the following inequalities hold:
\begin{align}
&\Bigl(\rho^m\vec u^{m+1}-\rho^{m-1}\vec u^m,~\vec u^{m+1}\Bigr)
 +\Bigl(\rho^{m-1}\left(\vec u^{m+1}-\vec u^m\right),~\vec u^{m+1}\Bigr)\nonumber\\
% &\quad = \Bigl(\rho^m\,\vec u^{m+1},~\vec u^{m+1}\Bigr)
%-\Bigl(\rho^{m-1}\,\vec u^{m},~\vec u^{m}\Bigr)+ \Bigl(\rho^{m-1}(\vec u^{m+1}-\vec u^m),
%~\vec u^{m+1}-\vec u^m\Bigr)\nonumber\\
&\qquad\qquad\geq \Bigl(\rho^m\,\vec u^{m+1},~\vec u^{m+1}\Bigr)
-\Bigl(\rho^{m-1}\,\vec u^{m},~\vec u^{m}\Bigr).
\label{eqn:energybound2}\\[0.5em]
&\Bigl(\partial_s\vec X^{m+1},~\partial_s(\vec X^{m+1}-\vec X^m)\Bigr)_{\Gamma^m}^h\geq|\Gamma^{m+1}| - |\Gamma^m|.
\label{eqn:energybound4}
\end{align}
Using \eqref{eqn:energybound2} and \eqref{eqn:energybound4} 
in \eqref{eqn:energybound1} and noting 
$x_r^{m+1}-x_l^{m+1} = |\Gamma_1^{m+1}|$, $
x_r^m-x_l^m =|\Gamma_1^m|$, we immediately obtain Eq.~\eqref{eqn:energybounds}. Replacing $m$ by $k$ in Eq.~\eqref{eqn:energybounds} and by summarising up for $k$ from $0$ to $m-1$, we obtain the global discrete
energy dissipation law \eqref{eqn:discreteenergylaw}.

\bibliographystyle{siamplain}
\bibliography{thebib}
\end{document}